\documentclass[11pt]{article}
 
\usepackage{mathtools}

\usepackage{graphicx}
\usepackage{amssymb}
\usepackage{amsmath}
\usepackage{upgreek}
\usepackage{multirow}
\usepackage[latin1]{inputenc}
\usepackage[labelfont=bf,labelsep=period,footnote size]{caption}
\usepackage[biblabel]{cite}
\usepackage{color}
\usepackage[breaklinks=true,colorlinks=true,linkcolor=blue,citecolor=blue,backref=none,pagebackref=false]{hyperref}
\usepackage{setspace}
\usepackage{epstopdf}

\usepackage{xcolor}
\definecolor{c1}{rgb}{0,0,1} 
\definecolor{c2}{rgb}{0.1,0.1,0.1} 
\definecolor{c3}{rgb}{0.3,0,0.9} 
\hypersetup{
    linkcolor= {c1}, 
    citecolor={c2}, 
    urlcolor={c3} 
}

\numberwithin{equation}{section}

\usepackage[auth-sc-lg,affil-sl]{authblk}

\pdfoutput=1
\usepackage{geometry}
\makeatletter
\renewcommand{\@biblabel}[1]{\quad#1.} 
\makeatother

\usepackage{setspace}
\usepackage{fancyhdr}
\fancypagestyle{empt}{
\fancyfoot[C]{\thepage}}
\newcommand{\diff}[2]{\frac{d #1}{d #2}}

\newcommand{\p}[0]{\mathbb P}

\newcommand{\A}[0]{\mathbf A}

\begin{document}

\title{Universal principles of cell population growth\\follow from local contact inhibition}

\author[1,\&]{Gregory J. Kimmel}
\author[1,\&]{Sadegh Marzban}
\author[2]{Mehdi Damaghi}
\author[3]{Arne Traulsen}
\author[1]{Alexander R. A. Anderson}
\author[1,*]{Jeffrey West}
\author[3,4,5,*]{Philipp M. Altrock}

\affil[1]{Department of Integrated Mathematical Oncology, Moffitt Cancer Center, Tampa, FL, USA}
\affil[2]{Stony Brook Cancer Center, Stony Brook School of Medicine, Stony Brook University, Stony Brook, NY, USA}
\affil[3]{Department of Theoretical Biology, Max Planck Institute for Evolutionary Biology, Ploen, Germany}
\affil[4]{Department of Hematology and Oncology, University Hospital Schleswig-Holstein, Kiel University, Germany}
\affil[ \&]{Equal contribution}
\affil[*]{Corresponding authors: jeffrey.west$@$moffitt.org, altrock@med2.uni-kiel.de}

\label{firstpage}

\maketitle

\section*{Abstract}
Cancer cell populations often exhibit remarkably similar growth laws despite their heterogeneity. Explanations of universal cell population growth remain partly unresolved to this day. Here, we present a growth-law unification by investigating the connection between microscopic assumptions and the expected contact inhibition, which leads to five classical tumor growth laws: exponential, radial growth, fractal growth, generalized logistic, and Gompertzian growth. All five can be seen as manifestations of a single microscopic model. Agent-based simulations substantiate our theory, and we can explain differences in growth curves in experimental data from {\em in vitro} cancer cell population growth. Thus, our framework offers a possible explanation for many mean-field laws used to empirically capture seemingly unrelated cancer or microbial growth dynamics. Our results highlight that the interplay between contact inhibition and other assumptions (e.g., well-mixed) can influence our quantitative understanding of how cancer cells grow and, in turn, how they may interact.

{\bf Keywords}: Population growth, cancer growth curves, cell motility, contact inhibition

\newpage

\onehalfspacing

\section*{Highlights}
\begin{itemize}
\item Five classical tumor growth laws emerge from a single microscopic model of contact inhibition.
\item The ratio of migration to proliferation determines which growth law governs population dynamics.
\item Agent-based simulations and in vitro experiments validate the theoretical predictions.
\item Gompertz growth arises at high density, explaining its failure at low cell counts.
\end{itemize}

\section{Introduction}

Cancer cell population dynamics often exhibit remarkably replicable, universal laws \cite{wodarz:book:2005,perez2020universal} despite their genetic, epigenetic, and phenotypic heterogeneity \cite{reiter:nrc:2019,hausser:nrc:2020}. The derivation of a universal growth law for tumors has been the subject of investigations for decades \cite{mombach:EPL:2002,bru:BJ:2003,drasdo:PB:2005,guiot:JTB:2003,rodriguez2013tumor,benzekry2014classical}. The winding search for a comprehensive mathematical characterization of tumor growth has belied two seemingly opposing concepts. First, a growth model should accurately describe the empirical data \cite{altrock:NRC:2015,brady:BMB:2019}. Second, a growth model should be based on and derived from a mechanistic biological framework \cite{cheng:pONE:2009,gerlee:CR:2013,West:PNAS:2019}. Cell population growth models with underlying mechanisms should be preferred because of their utility in comparative studies, model selection, and hypothesis generation. In cancer cell population dynamics, the underlying assumptions and macroscopic growth laws are essential ingredients for ongoing efforts to better understand adaptive therapy protocols \cite{strobl:AO:2023} and to enable treatment protocol personalization \cite{strobl:cellSys:2024}. 

The Gompertz growth model was originally employed as the best fit to the observed exponential decay of net cancer growth rates, and its widespread use can in part be explained by the use of double-logarithmic data transformation, leading to a linear regression procedure with difficult interpretation of the underlying biological variability \cite{gerlee:CR:2013}. Also, certain growth models are known to suffer from a lack of identifiable parameters given available data \cite{simpson:JTB:2022}. Attempts to mechanistically explain growth over time with time-dependent rates, such as Gompertz growth, remain largely unresolved. 

Gompertzian models find wide application, e.g.~in finance \cite{islam:IJF:2002}, immune cell population regrowth through niche occupation \cite{kimmel:procB:2021}, microbial dynamics \cite{zwietering:AEM:1990}, and tumors \cite{steel:CellPKC:1977,spratt:Cancer:1993}. Even without a biological mechanistic foundation, the use of Gompertzian growth models in cancer modeling indicates its predictive and descriptive potential \cite{benzekry2014classical}. Several studies have proposed a deeper understanding to reconcile the disconnect between descriptive power and mechanistic insight. Recent work by West \& Newton derives several related growth laws mechanistically by assuming gradual occupation of cellular microstates \cite{West:PNAS:2019,west2016evolutionary}. This approach indicates that population feedback between the populations' microscopic configurations can lead to macroscopic growth laws. 

In normal tissues, high cell density inhibits cellular proliferation upon reaching maximal capacity \cite{eagle1967growth,pavel2018contact}. In cancer cell populations, increased motility allows for continued movement after contact with neighbors \cite{ribatti2017revisited}. Loss of contact inhibition is common in solid tumors and increases invasion \cite{mayor2010keeping,mcclatchey2012contact}. Cell density is also influenced by cell migration and the size of the local neighborhood interaction \cite{Marzban24}. There is often assumed to be a dichotomy between migration (go) and proliferation (grow) \cite{hatzikirou2012go,gallaher:RSIF:2013,kimmel:CR:2020}. While the go-or-grow hypothesis remains controversial to some \cite{vittadello2020examining}, its logic lies in the observation that the cytoskeletal apparatus cannot perform both tasks at the same time \cite{syga:PLoSCB:2024} but might also be connected to DNA replication and genomic instability \cite{aguilera:nrg:2008,kimmel:CR:2020}. 

Contact inhibition can be broken down into two fundamental processes: contact inhibition of locomotion and contact-dependent growth inhibition. These mechanisms of growth inhibition are thus related to some degree of overcrowding, which is in contrast to slowing population growth due to undercrowding, known as the Allee effect \cite{allee:JEZ:1932,stephens:OIKOS:1999,gerlee:PCB:2022}. Contact inhibition initially referred to the inhibition of a cell's directional movement upon collisions with neighboring cells \cite{abercrombie1954observations}. Contact inhibition of locomotion (CIL) interacts simultaneously with attractive or repulsive chemotactic gradients\cite{lin2015interplay}, and has been implicated in cancer cell dissemination during metastasis\cite{stramer2017mechanisms}. CIL is related to contact inhibition of proliferation, often called contact-dependent growth inhibition (CDI). CDI does not depend on physical contact to slow proliferation, but can extend well beyond a cell's proximity\cite{dunn1984new, stramer2017mechanisms}.

Here, we present a unifying framework that leads to multiple classical growth laws based on the mechanism of local contact inhibition. Three major factors regulate contact inhibition and tumor cell density: proliferation, migration, and the size of the interaction region. Variations in these factors may lead to diverse macroscopic growth dynamics, but key properties of the dynamics can still be explained by simple biophysical mechanisms. The theory we introduce here considers five classical tumor growth laws: exponential, generalized logistic, Gompertz, radial, and fractal. Corroborated by agent-based simulations and statistical analysis of data from {\em in vitro} cancer cell population growth experiments, we show that these five growth laws can be captured by a single microscopic model under varied assumptions about contact inhibition. 


\section{Methods}\label{methods}

Here we provide a brief overview; detailed calculations and simulation results are presented in the Results section. 

\subsection*{Data and Code Availability}

Code and data used in this manuscript are publicly available at \\\url{https://github.com/MathOnco/Contact-Inhibition}. 

The agent-based model was implemented using the Hybrid Automata Library \cite{bravo:PCB:2020}. Data fitting was performed using Wolfram Mathematica (versions 12.0 and 14.1). Any additional information required to reanalyze the data reported in this paper is available from the lead contact upon request.

\subsection*{Experimental Model and Study Participant Details}

\subsubsection*{Cell lines}
For cancer cell culture experiments, we used the MCF-7 and MDA-MB-231 breast cancer cell lines, ovarian cancer cell lines OVCAR-3, OVCAR-4, A2780s, and TOV112D, and the PE9 murine cell line. All cell lines were acquired from American Type Culture Collection (ATCC, Manassas, VA, 2007--2010). MCF-7 cells are derived from a female patient; MDA-MB-231 cells are derived from a female patient; OVCA3, A2780s, and TOV112D are derived from female patients. Cells were maintained in RPMI 1640 (Life Technologies, Cat\# 11875-093) supplemented with 10\% fetal bovine serum (HyClone Laboratories) at 37$^\circ$C in a humidified incubator with 5\% CO$_2$. Cells were tested for mycoplasma contamination and authenticated using short tandem repeat DNA typing according to ATCC guidelines.

\subsection*{Methods Summary}

\subsubsection*{Immunofluorescence microscopy}
Cells were seeded on an 8-chamber microscopy slide overnight. They were then rinsed with phosphate-buffered saline (PBS), fixed in cold methanol:acetone (1:1) for 10 minutes, and further permeabilized with 0.5\% Triton X-100, then blocked with 5\% bovine serum albumin in PBS. Samples were incubated with LAMP2 rabbit primary antibody (1:100; ab218529, Abcam) and secondary Alexa-Fluor 488 anti-rabbit antibody (1:1000). Wheat germ agglutinin (WGA) was used to stain the membrane and DAPI for the nucleus. Coverslips were mounted using ProLong Gold Antifade Reagent (Life Technologies), and images were captured with a Leica TCS SP5 confocal microscope. Cell count measurements were performed with an oil-immersion 63$\times$ objective.

Initial confluence was controlled by seeding varying numbers of cells onto plates of identical surface area and quantified as the fraction of the plate area occupied by cells. All cells were cultured in a humidified incubator with 5\% CO$_2$ and 95\% relative humidity.

\subsubsection*{Analytical model derivation}
We defined a discrete, finite lattice with total lattice sites $l$, where each lattice site at position $\vec{q}_i$ is denoted by $\mathbf{A}(\vec{q}_i)$ and can take values of 0 (empty) or 1 (filled). The total number of cells on the lattice is $n = \sum_{i}^{l} \mathbf{A}(\vec{q}_i)$ where $n \in \{0, \ldots, l\}$. The neighborhood domain for cell $i$ is a set of coordinates denoted $\Omega_i$, with the number of elements $\omega_i$. Unless otherwise specified, we assumed constant, identical neighborhood sizes for all cells ($\omega_i = \omega$). The birth rate of a focal cell at position $\vec{q}_i$ is $\lambda$ if at least one neighboring site is empty, and 0 if the neighborhood is completely occupied. Death occurs at rate $\delta$ independent of neighborhood occupancy. The continuum approximation (equation 2.8) assumes that the domain is sufficiently large relative to the neighborhood size ($l \gg \omega$) such that boundary effects are negligible; cells near domain boundaries have reduced neighborhood sizes, but these represent a vanishing fraction of the population in the large-system limit. The Taylor expansions underlying the generalized logistic and Gompertz approximations assume well-defined moments of the birth neighborhood distribution. In our lattice model, this condition is automatically satisfied because $\omega_i$ is bounded by the finite neighborhood geometry (e.g., $\omega_i \leq 24$ for a second-order Moore neighborhood), ensuring that all moments exist and heavy-tailed distributions cannot arise.

\subsubsection*{Agent-based model simulations}
Stochastic simulations of individual-based models were performed using the Hybrid Automata Library software package \cite{bravo:PCB:2020}, which enables fast simulation of tumor growth models in spatial domains with real-time visualization. The model was implemented using Java JDK Version 8. Each stochastic simulation was initialized with a single cell placed in the center of a two-dimensional domain (typically 201$\times$201 lattice sites). At each time step, simulations loop over all focal cells in shuffled order. Each cell is checked for a migration event at a rate $m$; if migration occurs and there is a free lattice point within the migration neighborhood $\Phi$, the cell moves to a randomly chosen empty site. Next, the cell is checked for a birth event at rate $\lambda$; if a birth occurs and there is a free lattice point within the birth neighborhood $\Omega$, a daughter cell is placed in a randomly chosen empty neighboring site. If no birth occurs, the cell may undergo death at a rate $\delta$. Birth and death events are governed by independent rates; however, successful division additionally requires at least one empty neighboring site for daughter cell placement, whereas death occurs unconditionally. The time is then updated ($t_{i+1} = t_i + \Delta t$ where $\Delta t = \text{const.}$), and the process is repeated. Neighborhood configurations included Von Neumann ($\omega = 4$), Moore ($\omega = 8$), second-order Von Neumann ($\omega = 12$), and second-order Moore ($\omega = 24$) neighborhoods.

\subsubsection*{Growth law fitting}
We calibrated the Gompertz and generalized logistic growth laws to experimental cell count measurements using nonlinear regression. The generalized logistic model was $\dot{u} = \lambda u (1 - u^{\omega}) - \delta u$, and the Gompertz model was $\dot{u} = -\lambda \omega u \ln(u) - \delta u$. Fitting was performed using \textit{NonlinearModelFit} in Wolfram Mathematica (versions 12.0 and 14.1) with temporal confluency data from seven cell lines under variable initial seeding conditions. The birth neighborhood size $\omega$ obtained from fitting is an effective parameter obtained by nonlinear regression rather than a direct measurement of physical cell-to-cell contacts.

\subsection*{Quantification and Statistical Analysis}

Model comparison was performed using the Akaike Information Criterion (AIC). For each cell line and experimental condition, AIC scores were obtained for both Gompertz and generalized logistic models. Gompertz's AIC score was normalized by the generalized logistic's AIC score for comparison. This normalized score was plotted against initial confluence on a log-linear scale, and a linear fit was obtained using \textit{LinearModelFit} in Wolfram Mathematica. The adjusted $R^2$ for the aggregate trend was 0.24; however, all seven cell lines individually exhibited positive slopes, indicating that Gompertz fits improved with increasing initial confluence. Relationships between intrinsic growth rate (birth rate) $\lambda$ and birth neighborhood size $\omega$ were assessed using linear regression on log-transformed data. Statistical details, including $R^2_{\text{adj}}$ values, are reported in the figure legends (Figure 5) and main text. Sample sizes for agent-based simulations are indicated in figure legends. 



\section{Results}\label{results}

The quantities and parameters of the model are summarized in Table \ref{table:parameters}.   
\begin{table}[h]
\footnotesize
	\def\arraystretch{2}%
	\centering
	\begin{tabular}{|c|c|c|c|}
		\hline
		Parameter name & Symbol & Typical values & Range \\
		\hline
		Number of cells at time $t$ & $n(t)$ & $0,\dots,10^8$ & $[0,\infty)$ \\
		Total lattice sites & $l$ cells & $100,\dots,1000$ & $[0,\infty)$ \\
		Birth neighborhood domain/set for cell $i$ & $\Omega_i$ & -- & -- \\
		Birth neighborhood size for cell $i$ & $\omega_i$ & $4,\dots,8$  & $[0,24)$ \\
		Mean birth neighborhood & $\bar\omega$ & $4,\dots,8$  & $[0,24)$ \\
		Density-dependent birth rate at site $i$ with population size $n$ & $\lambda_{i,n}$ & -- & -- \\
		Density-dependent death rate at site $i$ with population size $n$ & $\delta_{i,n}$ & -- & -- \\
		Population average birth rate, death rate & $\lambda$, $\delta$ & -- & -- \\
		Cell migration rate (ABM) & $m$ & $0,\dots,1$ & $[0,\infty)$ \\
		Migration neighborhood set (ABM) & $\Phi_i$ & -- & -- \\
		Migration neighborhood size (ABM) & $\phi_i$ & 4,\dots,8 & $[0,\infty)$ \\
		\hline
	\end{tabular}
	\caption{Parameters and variables used in this work.}
	\label{table:parameters}
\end{table}

\begin{figure}[h]
\centering
\includegraphics[width=0.65\textwidth]{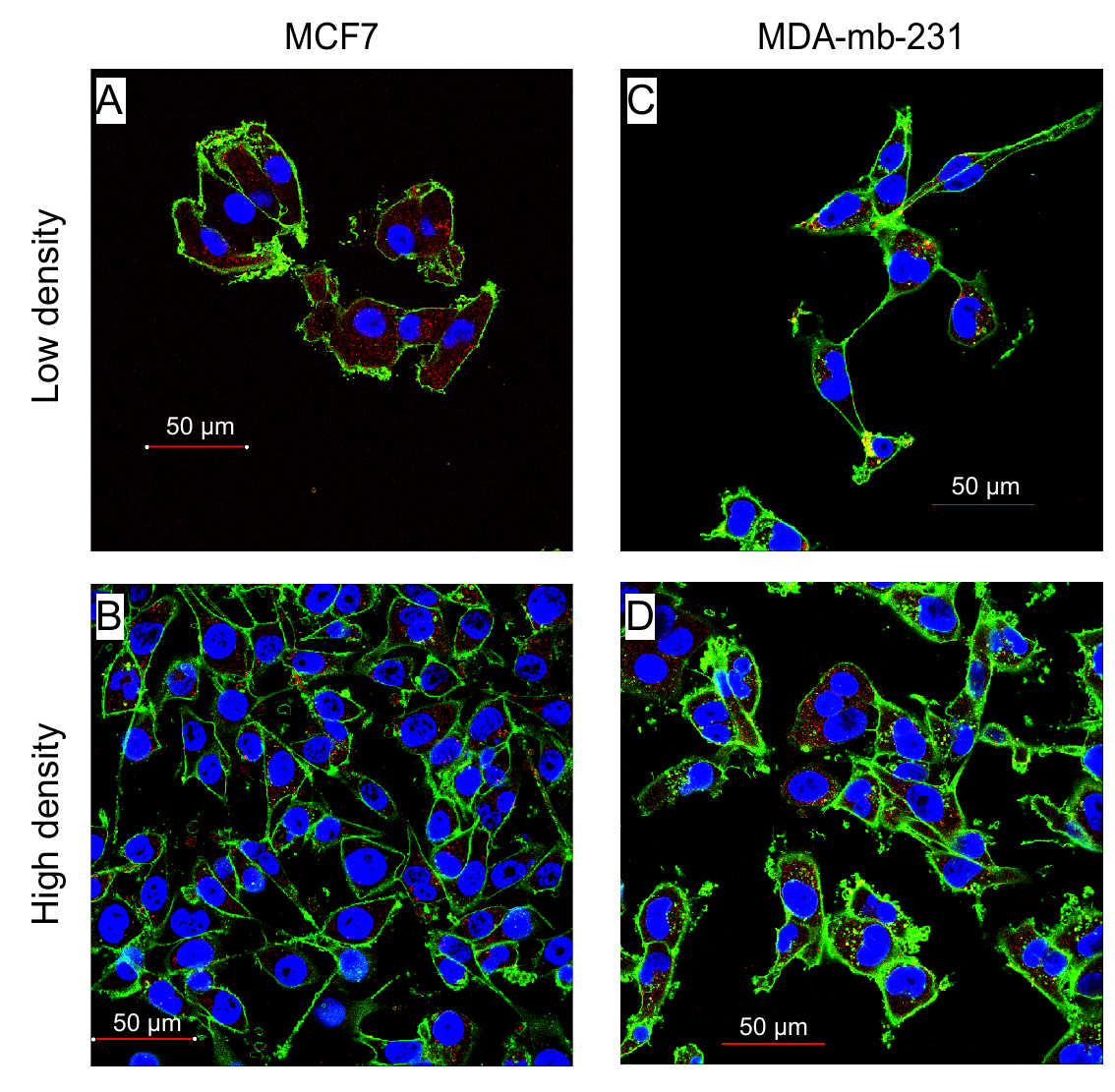}
\caption{
\textbf {Experimental observations of cancer cells distributed at low and high density.} \textbf{A, B:} Breast cancer cell line MCF-7 at low and high density. \textbf{C, D:} Breast cancer cell line MDA-mb-231  at low and high density.
}
\label{fig:Fig1}
\end{figure}

\begin{figure}[h]
\centering
\includegraphics[width=0.88\textwidth]{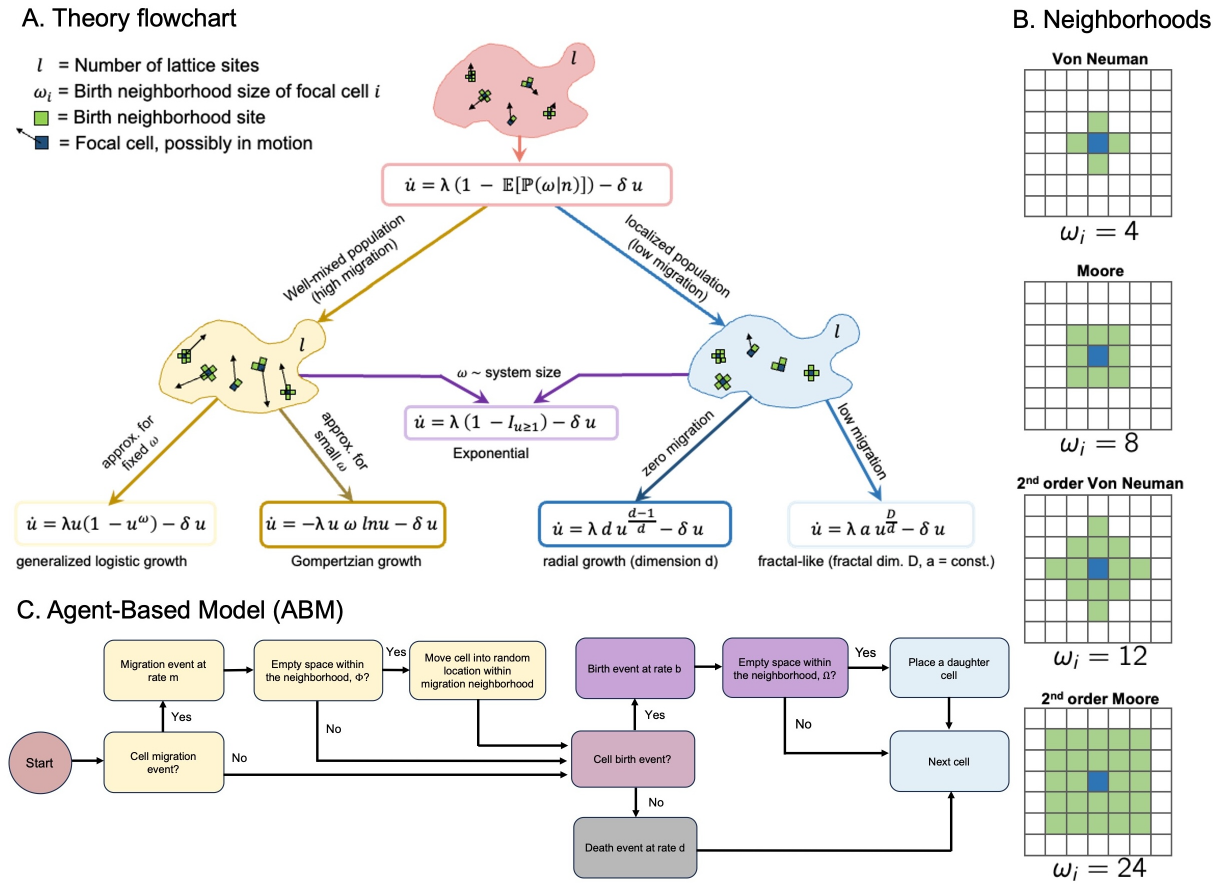}
\caption{
	\textbf{Overview of computational modeling and theoretical approaches.}
    \textbf{A:} Schematic of the theory hierarchy that describes the general model (red) and the limiting regimes we consider. We separate well-mixed population behavior (akin to high cell migration between birth events), and strict spatial localization, which leads to radial growth (akin to low migration between birth events).	 
	\textbf {B:} Protypical neighborhood paradigms for migration or birth for the individual-based model simulations.
	\textbf {C:} agent-based model (ABM) cell step flowchart. At each time step, the cell checks if a migration event occurs (at rate $m$),  and migrates if there is a free lattice point within the migration neighborhood.  Next, the cell checks if a birth event occurs,  and divides if there is a free lattice point to place the new daughter cell within the birth neighborhood.  If no free space is available, a birth does not occur. In the absence of birth, the cell may die (at rate $\delta$).
}
\label{setup}
\end{figure}

\subsection{Analytical framework}

Let us define a set of generalized coordinates $\vec q$ which define a discrete, finite lattice (with total lattice sites, $l$) where each lattice site at position $\vec q_i$ is denoted by $\A(\vec q_i)$ and can take on values of 0 (empty) or 1 (filled). Below, we will define an individual-based, on-lattice, birth-death-migration process that mimics contact inhibition of cell growth. The total number of $n$ cells on this lattice, $n = \sum_i^l \A(\vec q_i)$ where $n \in \{0,\dots,l\}$. 

Contact inhibition may occur within a localized region, or 'neighborhood,' around a cell, and is influenced by both cell migration Figure \ref{setup}A and the size of the neighborhood Figure \ref{setup}B. Commonly used neighborhoods for a two-dimensional lattice case are shown in Figure \ref{setup}B, but we can consider a neighborhood domain of any arbitrary size, dimension, and lattice connectivity. The neighborhood domain for the $i$\textsuperscript{th} cell is a set of coordinates which we denote as $\Omega_i$, with the number of elements $\omega_i$. For the rest of the manuscript, we assume constant, identical neighborhood sizes for all cells, $\omega_i = \omega$, unless we specify otherwise (e.g., for boundary sites).

First, we define the random variable $X_i \in \{0,1,\dots,\omega_i \}$ as the number of filled lattice sites neighboring cell $i$ that contain a living cell. Let $\p(X_i = x)$ be the probability of observing $x$ filled sites. We seek to derive the probability of observing $x$ neighbors given $n$ cells that arise as a result of a well-defined birth-death-migration process with a given set of assumptions. Formally, this expression is given by the following conditional probability:
\begin{equation} \label{eq: probability 1}
\p(X_i = x | n)
\end{equation}
and by definition:
\begin{align}
\sum_{x=0}^{\omega_i}\p(X_i = x | n)=1.
\end{align}

Each cell at location $\vec q_i$ divides at a rate $\lambda_i$ and dies at a rate $\delta_i$, and thus the number of cells is updated according to the following equation:
\begin{align}
	n(t+\Delta t) &= n(t) + \sum_{i=1}^n \Delta t\,\lambda_{i} - \sum_{i=1}^n \Delta t\,\delta_i \label{update_equation}
\end{align}

A  focal cell with a filled neighborhood is ``blocked'' from dividing. Thus, the birth rate of a focal cell at position $\vec q_i$ is given by
\begin{align}
\lambda_i=\begin{cases}
0: \text{if neighborhood is occupied, }x_i=\omega_i \text{, with likelihood } \p(X_i=\omega_i|n)\\
\lambda: \text{else}
\end{cases}
\end{align}
This statement for the birth rate of cell indexed $i$ assumes that a successful division requires only the availability of a single empty site in the local neighborhood. This assumption holds that contact inhibition primarily operates through spatial exclusion. As long as at least one neighboring site is available, division can proceed, whereas complete local crowding suppresses proliferation.

The probability of a focal cell's neighborhood containing {\em at least} one empty neighboring location is given by one minus the probability of a completely filled neighborhood, $1-\p(X_i = \omega_i | n)$. We also assume that the death rate is independent of neighborhood occupancy ($\delta_i = \delta$). Therefore, equation \eqref{update_equation} can be rewritten:
\begin{align}
	n(t+\Delta t) = n(t) + \Delta t\,\lambda \sum_{i=1}^n [1 - \p(\omega_i  | n)] - \Delta t\,\delta \,n.
\end{align}
For a small time step, as $\Delta t \rightarrow 0 $, we can rewrite this statement as a mean-field approximation for approximately continuous values of $n$ \cite{Allen:book:2003,gardiner:book:2004,kampen:book:1997}:
\begin{align}
	\diff{n}{t} &=   \lambda \sum_{i=1}^n [1 - \p(\omega_i  | n)] - \delta n. \label{n_equation}
\end{align}
Next, we take the expectation, $\mathbb E[\cdot]$ (see supplementary info for derivation), of the probability of a given focal cell being blocked, $\p(\omega  | n)$, and reduce the summation (where we drop the cell index $i$ for convenience):
\begin{align}
	\diff{ n}{t} &=   n\,\lambda \bigg(1 - \mathbb E[ \p(\omega  | n) ] \bigg) - n\,\delta.\label{main_equation}
\end{align}
This manuscript focuses on deriving the quantity inside the $\mathbb E[\cdot]$ for specific rulesets of the birth-death-migration process. For example, we will illustrate how the expectation varies with neighborhood size or assumptions about cell migration (see Figure \ref{setup}A). To connect to more familiar growth laws in the text below, we sometimes describe the change in the number of cells, $n$, or the change in the density $u = n/l$. For $u$, the equation (\ref{main_equation}) also holds:
\begin{equation}
    \diff{u}{t} =  u\,\lambda\bigg( 1 - \mathbb E[ \p(\omega | n)] \bigg) - u\,\delta, \label{MFT:1}
\end{equation}
This equation also follows from a diffusion approximation (see the supplement). The ordinary differential equation approximations for $n$ and $u$ assume that the domain is sufficiently large relative to the neighborhood size ($\gg \omega$) such that boundary effects are negligible and that expectations of products can be replaced by products of expectations. Cells near domain boundaries have reduced neighborhood sizes, but these represent a vanishing fraction of the population in the large-system limit. In the following sections, we derive the expected value of the probability of a blocked neighborhood, $\mathbb E[ \p(\omega | n)]$, and state its relation to familiar growth laws. 

In our analytical description, equation \eqref{MFT:1} presents a starting point from which multiple growth laws governed by contact inhibition at birth can be obtained. If the birth-neighborhood size is independent of $n$ and spatial distribution $\vec q$, then \eqref{MFT:1} reduces to that of exponential growth. The functional form of the growth law depends strongly on the behavior of the average number of neighborhood sites. In the following sections, we look at special forms of $\omega_n$ to discern some of the well-known growth laws that can emerge via contact inhibition. A schematic of the theory workflow, along with sufficient conditions, is shown in Figure \ref{setup}A.

\subsection{Radially expanding population}
A natural starting point is to assume a growing solid tumor with no migration. Using equation \eqref{MFT:1} as our point of departure, we assume the tumor is growing radially, where only cells on the outer surface with free space can divide. Let $r$ be the length scale for tumor cells, $R$ be the length scale of the tumor area, and $d$ be the dimension considered (typically $d=2$ for two-dimensional growth on a Petri dish). The number of tumor cells is then estimated by $n = (R/r)^d$. The number of cells not on the surface is given by $n_0 = (R/r - 1)^d$. Let us again define the random variable $X_i \in \{0,1,\dots,\omega_i \}$ as the number of filled lattice sites neighboring cell $i$ that contain a living cell, i.e., where an offspring of the focal cell cannot be placed. We now write out the probability of all sites occupied explicitly (and drop the index of the birth neighborhood size of the cell labeled with index $i$, as we assume that it is the same value for all cells: $\omega_i=\omega$),
\begin{equation}
    \p(X_i=\omega | n) = \begin{cases} 0 & \text{if } n = n_s, \\ 1 & \text{if } n = n_0. \end{cases}
\end{equation}
Defining surface cells $n_s$, we partition our expectation $n = n_s + n_0$, which leads to
\begin{equation}
    \mathbb E[ \p(\omega  | n)]=\frac{1}{n}  \left[\sum_{i=1}^{n_s}  \p(X_i=\omega | n) + \sum_{i=1}^{n_0}  \p(X_i=\omega | n) \right] = \frac{n_0}{n}.
\end{equation}
Using the relations between the length scales of tumor area versus population size leads to
\begin{equation}
    \frac{n_0}{n} = \left( 1 - \frac{r}{R} \right)^d = \left(1 - n^{-1/d} \right)^d \approx \left(1 - d n^{-1/d} \right).
\end{equation}
The last approximation is valid when $n \gg 1$. Here we use a first-order Taylor expansion, $(1-x)^d\approx1-d\,x$, with $x=n^{-1/d}$, which is valid in the large-population limit $n\gg1$. Biologically, this corresponds to tumors that are sufficiently large that surface effects act as a small correction to bulk growth. 
The deterministic growth law is then given by
\begin{equation}
\diff{n}{t} = \lambda d n^{\frac{d-1}{d}} - \delta n.
\end{equation}
With $d = 3$, we obtain the Von Bertalanffy model
\begin{equation}
\diff{{ n}}{t} = 3 \lambda n^{\frac{2}{3}} - \delta n.
\end{equation}
For radial growth on a Petri dish, we have $d = 2$ and obtain
\begin{equation}
\diff{{n}}{t} = 2 \lambda n^{\frac{1}{2}} - \delta n.
\end{equation}
Thus, from a general growth equation (\ref{MFT:1}) that includes the mechanism of movement and local contact inhibition, one can derive radial cell population growth laws based on a length-scale argument.

\subsection{Fractal growth}

The notion of self-similarity has been applied to investigate population growth; a radial expansion approximation might be poor at finer length scales. Fractals and dimensionality can significantly influence the governing laws that control growth. Although self-similarity of these shapes does not exist beyond a certain length scale (e.g., the radii of tumor cells is a natural cutoff), principles of fractal growth can be applied. This idea successfully showed the density falling off in silica particles described in \cite{orbach:science:1986}.

The total number of cells will approximately follow $n \propto R^d$ where $d$ is the actual dimension of the system and $R$ is the characteristic length scale of the growing population. In contrast, the surface dwelling cells $n_s \propto R^D$ where $D$ is the fractal dimension. It follows that $n_s \propto n^{D/d}$. This leads to a growth law
\begin{equation}
\diff{ n}{t} = \lambda \,a\,  n^{D/d} - \delta \,n,
\end{equation}
where we have introduced the constant $a$ related to the characteristic length of the tumor cell (and to the fractal dimension of the tumor). With non-fractal growth, $D = d-1$ and our model reduces to that of radial growth. However, fractal growth implies a dimension $d -1 < D \le d$. This implies:
\begin{equation}
    \p(\omega | n) = 1 - dn^{-1/d} \label{fractal_theory}
\end{equation}

To illustrate fractal growth, we ran stochastic simulations of an individual-based model (Figure \ref{setup}C) on a two-dimensional lattice, subject to varying neighborhoods (Figure \ref{setup}B), and migration rates using the Hybrid Automata Library \cite{bravo:PCB:2020}. Cells may migrate at rate $m$ within the migration neighborhood and divide at rate $b$ within the birth neighborhood. Importantly, the birth and migration neighborhoods may have different sizes.  Here, we do not consider spatially variant resource limitations, which have been shown to affect spheroid growth in previous work \cite{browning:eLife:2021}.
\begin{figure}\begin{center}
\includegraphics[width=\columnwidth]{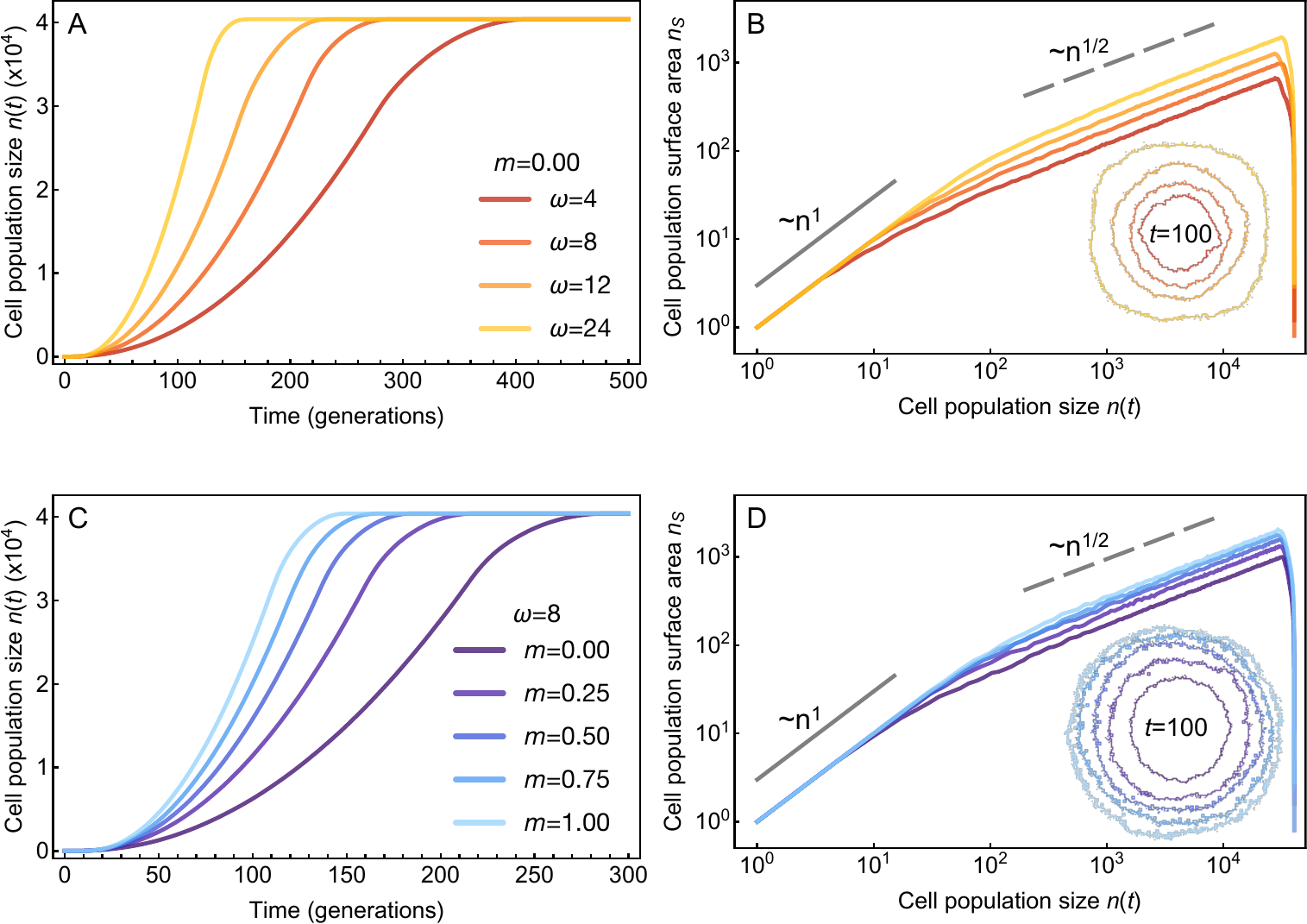}
\caption{
\textbf{Population growth dynamics of agent-based model simulations for the birth-death migration process.} 
Simulations are seeded with a single cell in the center of a $201 \times 201$ 2D domain, with periodic boundary conditions. Panels show results from $N=20$ stochastic realizations, with the bold line showing the mean value across replicates, and the shaded region shows one standard deviation.
\textbf{A:} Tumor growth in time and occupied lattice sites shown in panels A1, A2, for no migration ($m=0$), birth ($\lambda = 0.1$), death ($\delta = 0$), for varying birth neighborhood $\omega_b$. More growth is observed for larger birth neighborhood.
\textbf{B:} Tumor growth in time and occupied lattice sites shown in panels B1, B2, for high migration ($m=1$), birth ($\lambda = 0.1$), death ($\delta = 0$), for varying birth neighborhood $\omega_b$. 
\textbf{C, D:} Tumor growth (panel C) and fractal dimension (panel D) for varied migration rate ($m=0$-$1$). For this simulation only, the birth neighborhood is only the first-order Moore neighborhood ($\omega = 8$). Panels C2, D2 have an expanded $y$-axis.
}
\label{abm_results}
\end{center}\end{figure}

When migration is absent ($m = 0$), there is more potential growth for larger birth neighborhoods (Figure \ref{abm_results} A). However, when migration is high ($m = 1$; Figure \ref{abm_results} B), the birth neighborhood has no discernible effect on population growth. During high migration, cells are distributed randomly throughout the domain. On average, all cells have equal access to available space, regardless of birth neighborhood. In summary, the effects of the birth neighborhood are determined by the cells' migratory behavior.

The effects of contact inhibition are also mitigated by the increased migratory potential of tumor cells (Figure \ref{abm_results} C-D). As the migration rate, $m$, increases, growth potential (Figure \ref{abm_results} C) and surface area (Figure \ref{abm_results} D) also increase. Here, we assume a fixed migration neighborhood (1-Moore), but in the next section, we relax this assumption to explore a well-mixed assumption (i.e. a migration neighborhood which approximates the entire domain).

Simulations enable a numerical estimate of $\p$ as a function of the total population size, $n$. Figure \ref{fig:figure4}A depicts simulations initiated with a single cell at the center of the domain and calculated the fraction of the domain filled (n/l) and its corresponding numerical estimate of the likelihood $\p$, with no migration ($m=0$) for varied birth neighborhood size, $\omega$. Numerical estimates approach the theoretical prediction (eqn. \ref{fractal_theory}; long-dashed line shows $d=2$) for low $\omega$ values. 


\subsection{Well-mixed population growth}

Let us assume that the probability of observing $x$ filled sites is independent of the focal cell's location (e.g., on the expanding front of the growing population). This approximation is justified when cells move far between birth events. 
While migration is limited, correlations persist and lead to systematic deviations between the agent-based simulations and the corresponding mean-field growth laws, consistent with the need for correlation corrections described by \cite{baker:PRE:2010}. Our framework does not neglect correlations; rather, it identifies the migration regimes in which independence is a valid approximation and classical mean-field growth laws emerge.

Then, the probability of observing $x$ cells in a neighborhood of an arbitrary focal cell, given the population size is $n$, is  found by the number of ways of placing $x$ cells into the neighborhood and the remaining $n-x-1$ not in the neighborhood (recall there are only $n-1$ since the focal one has already been placed):
\begin{equation} \label{eq: hypergeo}
\p(x |n) = \frac{\binom{\omega_i}{x} \binom{l - \omega_i - 1}{n - x - 1}}{\binom{l-1}{n-1}}.
\end{equation}
We are interested in the probability that all neighboring sites are occupied, as this quantity enters the deterministic growth equation. For $l,n \gg \omega_i$ this leads to
\begin{equation}
\p(\omega_i |n) \approx \left(\frac{n-1}{l-1} \right)^{\omega_i} \approx \left(\frac{n}{l} \right)^{\omega_i},\label{well_mixed_theory}
\end{equation}
For the equation describing the dynamics of the expected fraction of occupied sites, $u=n/l$, based on equation (\ref{n_equation}), and assuming $\omega_i=\omega$ for all $i$, we obtain
\begin{equation} \label{eq: MFL well mixed}
\dot u = \frac{1}{l} \lambda \sum_{i=1}^n \left[ 1 - \left( \frac{n}{l} \right)^{\omega} \right ] - \delta u.
\end{equation}
Consider a Taylor expansion in $ \omega $ up to first order to evaluate the sum approximately. For this approximate expansion, we now describe two examples.  From a biological perspective, the Taylor expansion reflects the assumption that local variability in neighborhood crowding averages out at the population scale, so that growth is governed primarily by smooth, density-dependent regulation. This viewpoint aligns with a large body of cancer modeling work in which logistic and Gompertz laws are routinely used to describe tumor growth kinetics in leukemia, breast cancer, and metastatic lung disease \cite{benzekry2014classical,Vaghi,Tomelleri}.

First, consider a Taylor expansion around a typical value of birth neighborhood, $\bar\omega$. This value could be the population average, e.g., if the $\omega_i$ vary due to geometric constraints, $\bar\omega=(\omega_1+\omega_2+\dots+\omega_n)/n$. In addition, $\sigma_\omega^2=\sum\nolimits_{i=1}^{n}(\omega_i-\bar\omega)^2$, describes the respective standard deviation of the birth neighborhood across sites (which is typically very small). Taylor expansion up to the second order then gives
\begin{equation}
\dot u \approx \lambda u \left \{ 1 - u^{\bar\omega} \left[ 1 + \frac{1}{2} \sigma_\omega^2 (\ln u)^2 \right] \right \} - \delta u.
\end{equation}
We note that these Taylor expansions assume well-defined moments of the birth neighborhood distribution. In our lattice model, this condition is automatically satisfied because $\omega$ is bounded by the finite neighborhood geometry (e.g., $\omega = 24$ for a 2-Moore neighborhood), ensuring that all moments exist and heavy-tailed distributions cannot arise.

Provided we can neglect the standard deviation ($ 2 \gg (\sigma_\omega \ln u)^2$), we obtain the generalized logistic differential equation for the fraction of sites occupied by a cell
\begin{equation}\label{eq:genLogistic}
\dot u \approx \lambda\, u \left \{ 1 - u^{\bar\omega} \right \} - \delta \, u.
\end{equation}
This particular form is identical to Richards' differential equation, originally devised empirically as a flexible growth function that encompassed many growth laws \cite{richards:JEB:1959}. 

Note that we approximated the neighborhood size by its mean, which is valid when variability in local neighborhood occupancy is small or nonexistent. This approximation does not hold in very low-density regimes where fluctuations may dominate. 

Second, consider a Taylor expansion near 0. This na\"ive case could arise if sites with $\omega_i=0$ dominate the population's ``average" neighborhood size. We obtain (to second order)
\begin{equation}
\dot u \approx -\lambda u \left [ \bar\omega \ln u + \frac{1}{2} \left( \sigma_\omega^2 + \bar\omega^2 \right) (\ln u)^2 \right ] - \delta u.
\end{equation}
We can neglect the higher-order term if the following condition is met:
\begin{equation} \label{Gompertz condition}
\frac{2\bar\omega}{\sigma_\omega^2 + \bar\omega^2} \gg |\ln u|.
\end{equation}
This further approximation thus leads to
\begin{equation} \label{eq:GompertzGrowth01}
\dot u \approx -\lambda\,\bar\omega\,u \ln u  - \delta u,
\end{equation}
which we recognize as a form of Gompertz's law. The ``Gompertz condition," given by \eqref{Gompertz condition}, provides a mathematical justification for why Gompertz growth law descriptions are notoriously poor (and often incorrect) when the cell count is small. This condition is derived within the same weak-dependence regime discussed above, in which migration suppresses spatial correlations and neighboring site occupancies are approximately independent. Modifications to alleviate this issue have been introduced; for example, the Gompertz-Exponential model proposed by Wheldon, which assumes that growth is initially exponential before switching to a Gompertz law \cite{wheldon:book:1988}. If we stipulate that Gompertz emerges via contact inhibition, then a necessary condition for this law to be valid is that $u$ is not too small. This occurs through the $\ln u$ term in \eqref{Gompertz condition}, since a sufficient condition is that $u \approx 1$ (i.e., $n\approx l$) or that the mean number of neighbor sites $(\bar\omega)$ is very small.

Interestingly, one does not recover logistic growth by assuming that every tumor cell can place offspring at each site. Relation \eqref{eq: hypergeo} shows that when $x = \omega$ and $\omega = l -1$, the conditional probability is given by
\begin{equation}
\p(X_i=\omega | n) = \frac{\binom{0}{n-l}}{\binom{l-1}{n-1}} = \begin{cases} 0 & \text{if } n < l \\ 1 & \text{if } n \ge l \end{cases} := I_{n \ge l}.
\end{equation}
Here, $I_{n \ge l}$ is an indicator function which is 1 when satisfied and 0 otherwise. Using the reasoning from above leads to the deterministic growth law
\begin{equation}
\dot u = \lambda \, u \left(1 - I_{u \ge 1} \right) - \delta \,u,
\end{equation}
which implies exponential growth until saturation if $\lambda > \delta$, which differs from the logistic growth approximation, equation (\ref{eq:genLogistic}). 

\begin{figure}\begin{center}
\includegraphics[width=\columnwidth]{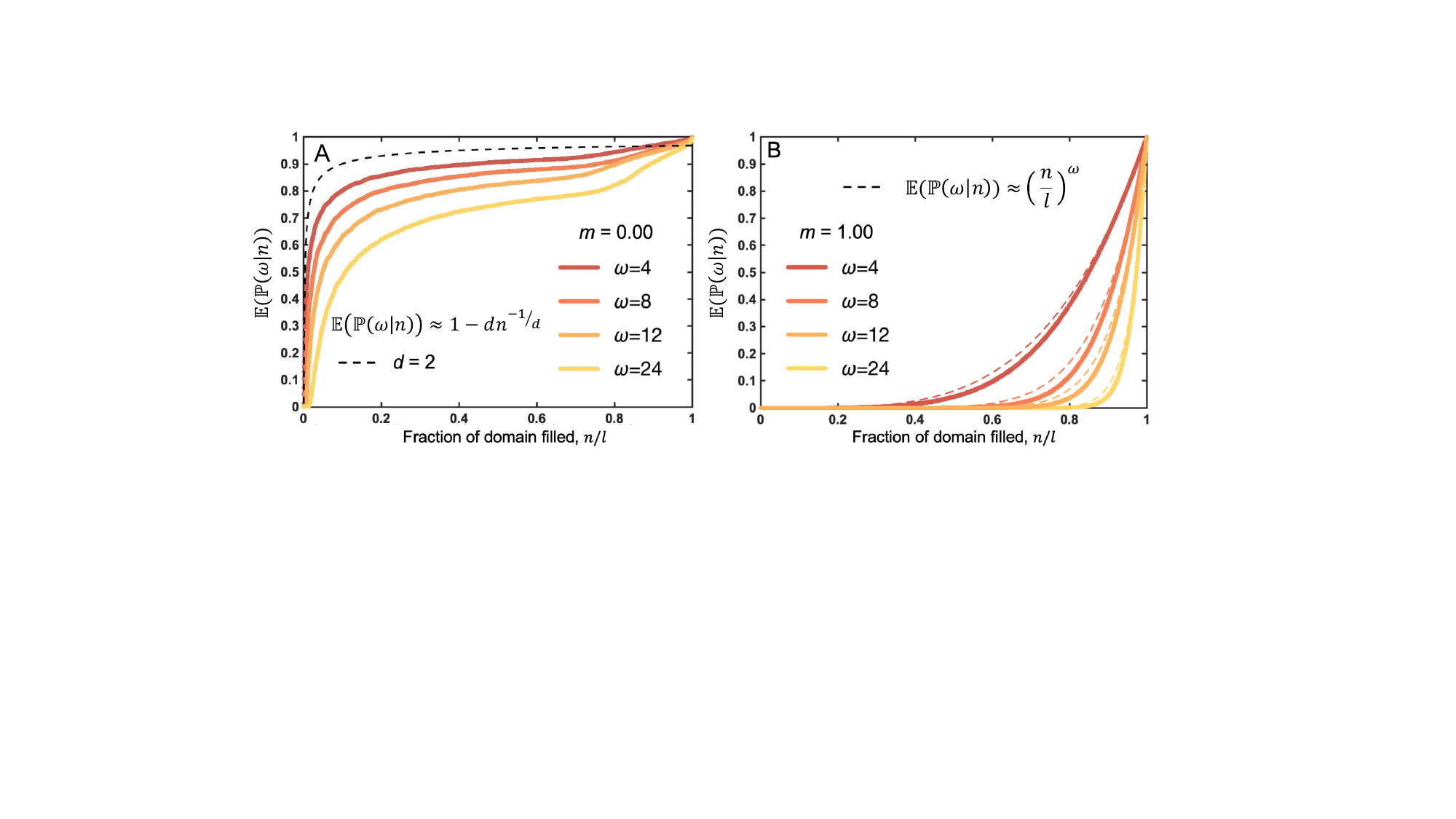}
\caption{
\textbf{The probability of observing $x$ neighbors given a total population of $n$ cells.} Mean of N stochastic realizations of an agent-based model seeded from a single cell in the center of the domain. The likelihood function, $\p$, is estimated by summing the total number of cells with a completely filled neighborhood ($x_i=\omega_i$) divided by total lattice size, $l=4096$. 
 \textbf{A:} No migration, varied birth neighborhood size. Birth rate is $\lambda = 0.1$, and migration is $m=0$. Long-dashed line shows theoretical prediction (shown for $d = 2$; see eqn. \ref{fractal_theory}); short-dashed line shows $d=3$. The simulations approach the upper theoretical prediction as $\omega\rightarrow0$. \textbf{B:} High migration, varied birth neighborhood size. The birth rate is $\lambda = 0.1$ and the migration rate value is $m=1$, with the neighborhood size of  $l$. Our theoretical prediction (Eq. (\ref{well_mixed_theory})) aligns well with the simulation data in the limit when $l, n \gg \omega$.}
\label{fig:figure4}
\end{center}\end{figure}

Simulations provide a numerical estimate of $\p$ as a function of the total population size, $n$, in Figure \ref{fig:figure4} B for a well-mixed scenario (high-migration rate; $m=1$). Theoretical predictions (Eq. (\ref{well_mixed_theory})) align well with the simulation data in the limit when $l, n \gg \omega$. The dashed line converges with the data at higher $n$ when $\omega$ is large.

In summary, our agent-based simulations confirm theoretical predictions for the probability of observing $x$ neighbors given a total population of $n$  cells in two scenarios: zero (or low) migration (Figure \ref{fig:figure4} A and high migration in Figure \ref{fig:figure4} B. The summary of assumptions is presented in Table \ref{table:theoretical_predictions}, along with the conditions for validity of the approximations used throughout the manuscript. The validity conditions specify the regimes where each growth law applies: radial growth holds for small populations relative to the neighborhood size; fractal-like growth emerges once populations are sufficiently large; exponential growth requires that cells can always find space to divide (birth neighborhoods spanning nearly all sites); generalized logistic growth applies when birth neighborhood sizes are tightly distributed around their mean; and Gompertz growth arises when birth neighborhoods are very small on average and the population is not at extreme sizes.

\subsection{Integrating theory with {\em in vitro} and {\em in silico} data}

We fit the growth laws derived in the previous sections to experimental cell count measurements. For cells growing in a two-dimensional domain, we expected the average number of neighbor sites to range from 1 to 8. This parameter could be slightly lower overall due to a reduction near the domain's boundary. Still, this reductive effect should be small for sufficiently large domains relative to cell size. We calibrated the Gompertz and generalized logistic growth laws using {\it NonlinearModelFit} (Wolfram Mathematica, v. 12.0 or 14.1) with temporal data from seven cell lines with variable initial seeding conditions in two-dimensional in vitro cell culture.

\begin{figure}
\begin{center}
\includegraphics[width=0.95\textwidth]{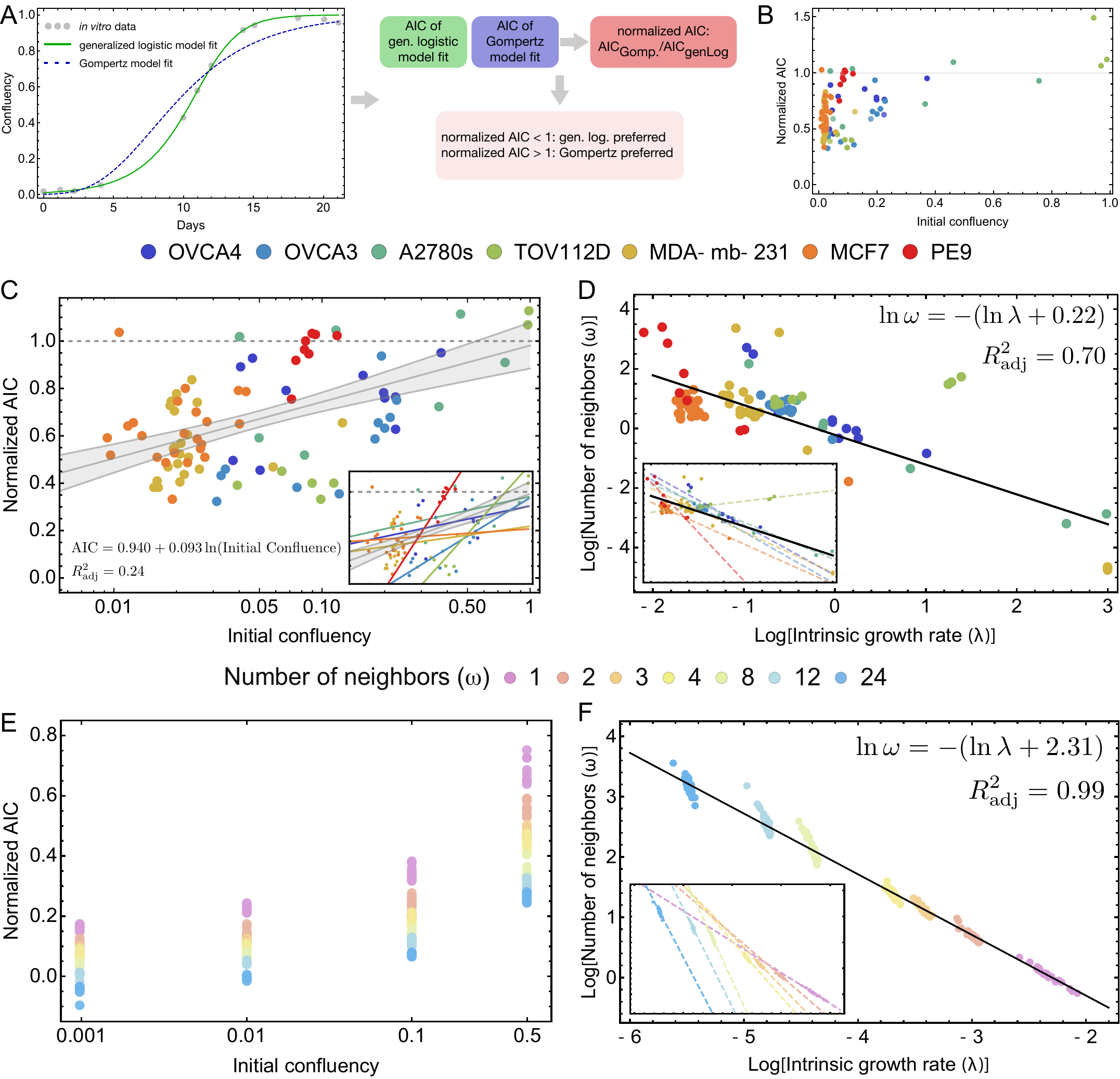}
\caption{\textbf{Assessing model-fits for the different cell lines (given above) for {\em in vitro} and {\em in silico} ABM experiments, using an Akaike Information Criterion (AIC).} 
\textbf A: Workflow for the {bf normalized AIC} value for a given experimental setup consisting of an initial confluency and a cell line.  
\textbf B: Linear relationship of initial confluency and normalized AIC. 
\textbf C: Confluency vs. AIC\textsubscript{Gompertz}/AIC\textsubscript{Gen. Log.} score, along with a best-fit line that shows the general improvement of Gompertz fits as the initial confluency increases. 
\textbf D: Intrinsic growth rate vs.~number of neighbors $(\lambda,\omega$) on a log-log scale. Different experimental setups (initial confluence and cell line) are colored, and a best-fit line is determined. In Supplementary Figures S1 and S2, we show individual fits using generalized logistic and Gompertz growth laws, respectively. 
\textbf E: Confluency vs. AIC\textsubscript{Gompertz}/AIC\textsubscript{Gen. Log.} score. 
\textbf F: ABM Intrinsic growth rate vs. Number of neighbors $(\lambda,\omega$) on a log-log scale. Different birth neighborhood sizes are color-coded, and a best-fit line is drawn.}
\label{fig:experiment_results}
\end{center}\end{figure}

A prediction obtained from the derivation of Gompertz growth via contact inhibition was that the approximation to the mean-field growth law would be poor if the initial confluence were low. This effect has been noted both phenomenologically \cite{wheldon:book:1988} and experimentally \cite{steel:CellPKC:1977}. To test this effect in our framework, we compared the relative goodness-of-fit between Gompertz and generalized logistic as a function of initial confluence for each cell line and each experimental setup (amount of initial confluence). An Akaike Information Criterion (AIC) score was obtained for both models (\ref{fig:experiment_results}A), whereby Gompertz's AIC score was normalized by the generalized logistic's AIC score for comparison. We plotted this normalized score against the initial confluence (\ref{fig:experiment_results}B, C) and fit a line using {\it LinearModelFit} (Wolfram Mathematica versions 12.0 and 14.1). The resulting fit and cell line data are shown in Figure \ref{fig:experiment_results}C. The dashed line at 1.0 is meant to guide the eye; values above this imply that Gompertz is an improvement over generalized logistic. This is not necessarily to say the respective fits are better, but rather that generalized logistic provides no information gain, as it carries an extra parameter. With an $R_{\text{adj}}^2 = 0.24$, the straight-line fit is not strong, but a general upward trend was observed as a function of initial confluence. We plotted each cell line's fit (Figure \ref{fig:experiment_results}A inset). The positive slope for each cell line indicates that Gompertz generally improves with higher initial confluence experiments.

We acknowledge that the aggregate trend in Figure \ref{fig:experiment_results}C is somewhat weak, reflecting substantial biological variability across cell lines and experimental conditions. However, the consistency of the effect is more apparent when examining each cell line individually (Figure \ref{fig:experiment_results}C, inset): all seven cell lines exhibit a positive slope, indicating that Gompertz fits improve relative to generalized logistic fits as initial confluence increases. This unanimous directionality across cell lines supports the theoretical prediction that the Gompertz approximation becomes valid at higher confluence, even though the aggregate relationship is noisy.

Next, we fit the \emph{in silico} experiments to the growth laws derived in the previous sections. For cells growing in a sufficiently large Petri dish, we expect that, on average, the number of neighbor sites will be only weakly impacted by the boundary. This effect is slightly diminished by the reduction near the boundary, but it will be small for sufficiently large systems relative to the cell size.
The derivation of Eq.~\eqref{eq: MFL well mixed} assumes that the probability of every site being occupied is independent of site location. A natural rule that justifies this assumption is that cell movement occurs much faster than the cell cycle. We further assume an inverse relationship exists between $\lambda$ and $\omega$ of the form $C = \lambda \omega$. This assumption is based on the results of the fitting routine applied to the aggregate experimental data (Figure \ref{fig:experiment_results}D).
Inferred intrinsic growth rate (birth rate) vs. the inferred birth neighborhood plotted on a log-log scale. Data points are colored by experimental condition (cell line), and the best-fit line to the correlation is shown.

In the agent-based model, migration is implemented as a simple random walk with exclusion, and it regulates spatial correlations: low migration preserves geometric constraints on proliferation, whereas high migration suppresses correlations, enabling a mean-field description of neighborhood occupancy. We ran the ABM with $\lambda \omega = 0.1$ and $\delta = 0.001$, applying the same fitting procedure described to handle the {\em in vitro} data. We then computed analogous statistics comparing goodness-of-fit, comparing growth rate vs.~number of neighbors (Figure \ref{fig:experiment_results}E, F). We found, as predicted, that higher initial confluency improves Gompertz's goodness-of-fit (relative to generalized logistic) and that the fits also improve as $\omega \to 0$. However, the Gompertz-based fitting procedure never reached the same goodness-of-fit as generalized logistic (compare this to Figure \ref{fig:experiment_results}C). These differences in goodness-of-fit likely stem from the difficulty of providing a natural rule to drive an ABM framework at the single-cell level using the birth neighborhood. Despite this discrepancy, the general trend of fit improvement with increasing initial confluence and decreasing number of neighbors is consistent with the theoretical model. For examples see Supplemental Figure S4.

We then looked at the relationship between growth rate and number of neighbors; using linear regression, we recaptured the growth law given to the ABM, by noting that if we take the log of both sides of $\lambda \omega = 0.1$ and solve, we obtain $\log \omega = -(\ln \lambda + 2.30)$) (see Figure \ref{fig:experiment_results}F). However, we also observed an interesting trend within the groups defined by fixed birth neighborhood $\omega$. Instead of looking at the aggregate, if we consider each value of $\omega$ individually and fit a linear regression, the slopes approach -1, which implies an inverse-power law. Both the cell lines and \emph{in silico} ABM simulations show a change in slope of the population surface growth, trending towards an inverse power law as $\omega$ decreases. We discuss this relationship in more detail in Section \ref{discussion}.

The birth neighborhood size $\omega$ obtained from fitting is an effective parameter obtained by nonlinear regression of the generalized logistic or Gompertz growth laws to experimental confluency data. In that sense,  it may not be a direct measurement of physical cell-to-cell contacts. The values fitted to {\em in vitro} data span approximately three orders of magnitude, whereas the values fitted to the ABM data closely recapitulate the value used in these simulations. The broad range of values from the cell line data arises because the effective $\omega$ absorbs multiple biological factors that modulate contact inhibition beyond simple neighbor counting, including cell shape, motility, adhesion properties, and deviations from the well-mixed assumption. This range is also consistent with the known biological variation in cancer cell line proliferation rates: doubling times in the NCI-0 panel range from 17 to 80 hours \cite{polymenis:CD:2017}, and individual cell lines can exhibit doubling times from as short as 12 hours to several days depending on culture conditions. Since our theory predicts an inverse relationship between $\lambda$ and $\omega$ (Figure \ref{fig:experiment_results}D), roughly one order of magnitude variation in birth rate produces compensating variation in the fitted value for $\omega$. In particular, the theory permits situations where geometric or crowding effects severely restrict a cell's ability to place offspring, even when physical neighbors are few. Thus, the fitted values could be understood as a phenomenological parameter capturing the aggregate strength of contact inhibition and reflecting the known heterogeneity of the cell lines' birth rates.

\subsection{Competition and heterogenity}

\begin{figure}
\begin{center}
\includegraphics[width=0.9\textwidth]{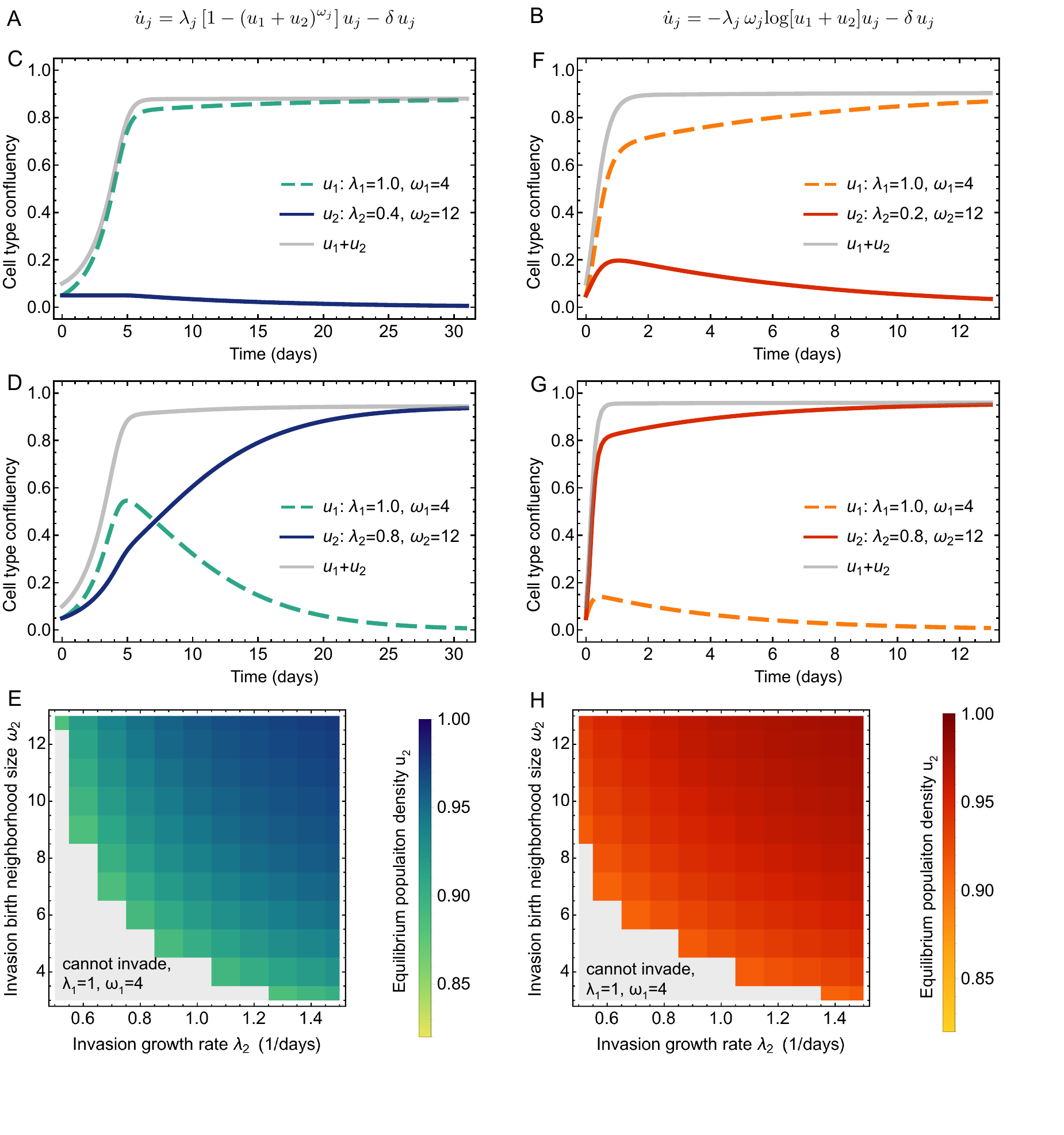}
\caption{\textbf{Ad hoc models of competition among two subpopulations with differing growth rates and birth. neighborhoods.} 
\textbf A: Generalized logistic growth of two subpopulations. 
\textbf B: Gompertz growth of two subpopulations. 
\textbf C, \textbf D, \textbf E: Extinction and takeover (invasion) of a disadvantaged subpopulation with a higher birth neighborhood (generalized logistic). 
\textbf F, \textbf G, \textbf H: Extinction and takeover (invasion) of a disadvantaged subpopulation with a higher birth neighborhood (Gompertz). 
Parameter values are given in each panel. Note that the indices $i$, $j$ here refer to the entire subpopulation and not to individual cells; all cells within a subpopulation described by confluency $u_i$ have the same growth rate $\lambda_i$ and birth neighborhood $\omega_i$. 
 }
\label{fig:heterogeneity}
\end{center}\end{figure}

Motivated by our numerical and statistical findings, we broadly asked if and how heterogeneity in growth rate and birth neighborhood size would impact the dynamics. To this end, we implemented an {\em ad hoc} superposition of growth models of two populations described by either Eq.~(\ref{eq:genLogistic}) or Eq.~(\ref{eq:GompertzGrowth01}), whereby the two simultaneously growing populations differ in their intrinsic growth rates (birth rates) and the sizes of the birth neighborhoods. We assumed that within each subpopulation, all cells have the same parameter values. These dynamics are shown in Figure \ref{fig:heterogeneity}. There, we show the dynamics of two competing subpopulations either governed by a generalized logistic growth law (Figure \ref{fig:heterogeneity}A) or a Gompertzian growth law (Figure \ref{fig:heterogeneity}B). While the overall confluency of the total population follows a consistent monotonic growth curve, the two subpopulations can exhibit a rich set of behaviors, as the ability to advance offspring into a larger set of spatial sites (larger birth neighborhoods) can confer a selection advantage. 

To better understand competition among subtypes of cells with different birth neighborhoods, we specifically asked under which birth neighborhood differentials a cell type with a lower birth rate can take over. Figure \ref{fig:heterogeneity} panels C and D show two example cases of competitive expansion under the generalized logistic growth law. If the birth rate of the second population is sufficiently large, the respective subpopulation cannot overtake the system, even with a large birth neighborhood. However, for a certain critical value of a higher yet still disadvantaged birth rate, the same subpopulation can invade. Overall, Figure \ref{fig:heterogeneity} D summarizes this behavior: above which value of growth rate a disadvantaged (or favored) subpopulation can invade due to a higher (or even with a lower) birth neighborhood. 

For the case of two competing subpopulations under a Gompertz growth law (Figure \ref{fig:heterogeneity}B), we observe a similar behavior (Figure \ref{fig:heterogeneity} F, G, H). Additionally, if we compare the invasion landscapes shown in Figures \ref{fig:heterogeneity} E, H, it is slightly easier for a growth-disadvantaged subpopulation to invade due to a higher birth neighborhood. This interpretation has to be taken with a grain of salt because, as shown above, the Gompertz growth law describes population dynamics for sufficiently high confluency or small birth neighborhoods. Taken together, these results from an ad hoc competition model of two subpopulations with potentially drastic differences show how costly a reduction in growth rate can be. This cost can be offset by greater flexibility in placing daughter cells within a larger birth neighborhood. 

%


\section{Discussion}\label{discussion}

The emergence of multiple well-known laws governing cancer cell population growth from a single individual-based process via contact inhibition is striking. Given the prominence of cell-based models in cancer biology \cite{rejniak:review:2011,metzcar2019review,fadai:PRSA:2019,chamseddine2020hybrid}, it is crucial to consider how model construction affects population growth dynamics. Suppose we stipulate that cancer cells' primary impediment to division is due to their inability to place offspring in a nearby location. In that case, the mean-field (deterministic) law determining tumor growth is not necessarily unique. Multiple growth laws can adequately describe \emph{in vitro} cancer growth, as they approximate mean-field dynamics in different limits. Multiple laws could be valid at the intersection if these approximations hold simultaneously. We note that previous work suggests that short time scales typically used in cell proliferation assays (e.g., 24 hours) may be sufficient in estimating growth rate parameters (e.g., $\lambda$); however, these time scales cannot reliably estimate the effects of contact inhibition at high confluency \cite{warne2017optimal}.

Correlation-correction approaches,  such as those of Baker \& Simpson \cite{baker:PRE:2010},  begin from microscopic birth-death-movement dynamics and derive continuum descriptions that explicitly incorporate spatial correlations; our framework, instead, asks which macroscopic growth laws emerge directly from microscopic interaction rules and spatial mechanisms. Our work complements earlier studies, such as Mombach et al.~\cite{mombach:EPL:2002}, showing that not only the form of inhibition but also spatial organization and migration can select among distinct growth regimes.

The connection between Gompertz and generalized logistic is well known and requires $\omega \to 0$ and $r \to r_0/\omega$, which implies that $r \omega = r_0$ in the limit of small $\omega$. We thus fit the cell line data to this power law and obtained an adjusted $R^2$ of 0.7. In Figure \ref{fig:experiment_results}D, it is clear that cell lines with a larger birth neighborhood size do not necessarily follow the same inverse-power law, but as $\omega$ gets smaller, the power seems to approach $-1$. Additionally, including the previously neglected higher-order terms in Generalized Logistic and Gompertz (Table \ref{table:theoretical_predictions}; right column) may lead to more flexible, modified versions of these growth laws.
\begin{table}[h]
\footnotesize
	\def\arraystretch{3}%
	\centering
	\begin{tabular}{|l|c|c|c|c|}
	\hline
		{\bf Name} & {\bf Growth dynamics}, $\dot{u}$  & $\mathbb E[ \p(\omega  | n)]$ & {\bf Migration} & {\bf Valid when...} \\ 
	\hline
           \hline
		Radial & $\lambda\, d \, u^{\frac{d-1}{d}}-\delta u$ & $1 - n^{-1}$ & None & $n \ll l$ \\
           \hline
            	Fractal-like & $\lambda\,a\, u^{\frac{D}{d}}-\delta u$ &  $1- dn^{-1/d}$ & Low & $n \gg 1$\\
           \hline
            	Exponential & $\lambda u (1 - I_{u\geq 1})-\delta u$ & $0 \text{ if } n < l$; $1 \text{ if } n \ge l$ & - & $\omega = l-1$ \\
           \hline
            	Gen. logistic & $\lambda u (1 - u ^{\bar\omega})-\delta u$ & $\left(\frac{n}{l} \right)^\omega$ & High & $\omega \approx \bar\omega$; $ 2 \gg (\sigma_\omega \ln u)^2$\\
           \hline
            	Gompertz & $-\lambda \, \bar\omega \, u \log (u) - \delta u$ & $\left(\frac{n}{l} \right)^\omega$  & High & $\bar \omega \ll 1$; $\frac{2\bar\omega}{\sigma_\omega^2 + \bar\omega^2} \gg |\ln u|$  \\
	\hline
	\end{tabular}
	\caption{
	{\bf Summary of theoretical predictions.} $D$ is the fractal dimension, and $a$ is a constant related to the characteristic length of the tumor cell. The quantity $\bar\omega$ represents a cell population average or typical value of birth neighborhood size that does not have to be an integer and can be close to 0 (representing a case where the cell population runs out of birth neighborhoods due to geometric or crowding effects). The quantity $\sigma_\omega$ is the respective standard deviation. Both $\bar\omega$ and $\sigma_\omega$ are static quantities. \\
	The validity conditions specify the regimes where each growth law applies: radial growth holds for small populations relative to the neighborhood size; fractal-like growth emerges once populations are sufficiently large; exponential growth requires that cells can always find space to divide (birth neighborhoods spanning nearly all sites); generalized logistic growth applies when birth neighborhood sizes are tightly distributed around their mean; and Gompertz growth arises when birth neighborhoods are very small on average and the population is not at extreme sizes.
	}
	\label{table:theoretical_predictions}
\end{table}

Why do cell lines have different relationships between (average) birth neighborhood size $\omega$ and net growth rate $\lambda$? Assuming a cell requires access to an open area for division, a larger surface area would enable it to sense open space. Thus, a natural prediction emerges from this theory: cells with a large $\omega$ are more likely to be irregularly shaped. Elongated cell morphology can increase the packing density of cells \cite{pichugin:PNAS:2022}, effectively increasing the number of cells per unit area, thus increasing any effect of contact inhibition (corresponding to an increase of $\omega$ in our framework). Similarly, previous work has shown that cells that switch from high polarity (e.g., elongated) at low confluency to low polarity at high confluency are associated with higher growth rates\cite{castro2003gompertzian}. Malleable cell shape could easily be related to a time-dependent $\omega$, or cell adhesion could impact the number of allowed neighbor-sites \cite{anderson:MMB:2005}. Birth neighborhood size also affects tumor-scale morphology, with larger neighborhoods tending to yield more diffuse tumor boundaries \cite{Marzban24}. While the analysis above focused on cancer cell lines, it applies to any migratory, proliferative cell line. Non-cancerous cell lines would typically be associated with lower growth rates (e.g, $\lambda$) and thus predicted to have a large birth neighborhood size, $\omega$. Existing data primarily focus on cancer cell population growth under various conditions (genetic changes, etc.), and data on normal cell behavior are difficult to find.  Herein, we consider competition for space as a primary driver of emergent growth dynamics, but experimental conditions (especially acidity, oxygenation, and resource availability) also affect the growth dynamics of cancer cell lines \cite{Freischel:SR:2021}.  We speculate that typical epithelial cells would have birth neighborhoods of $\omega=2$ or $4$. At the same time, proliferation and death rates could vary substantially depending on the organ of interest (colon, skin, brain, or bone marrow).

Cell-to-cell signaling could be incorporated via a local dependency on the birth-neighborhood $\omega$. The related molecular factors are often secreted in a paracrine manner and can promote or inhibit a cell's ability to proliferate. Including such factors can lead to a spatially-dependent $\omega$ or $\lambda$ (see Supplemental Figure S3). Recent studies have begun to explore how global growth patterns emerge from local cellular interactions \cite{fadai:PRSA:2020,gerlee:PCB:2022}, opening new avenues for understanding population growth dynamics governed by general principles. Tissue- or tumor-specific patterns emerge from the magnitude of interactions, such as those involving inhibitory or growth-factor signaling. 

Fractal or radial growth demonstrates how the proper deterministic growth law can change over time. All cells are effectively surface-dwelling at low cell surface area (or low volume, in three dimensions). Hence, the dynamics of all radial- or fractal-growing tumors are initially exponential. The shift from exponential to fractal growth depends on the birth neighborhood size. Unless the birth neighborhood size equals all sites ($\omega_i = l$), the tumor will transition from exponential to fractal-like growth, most easily seen on a log-log scale. As the tumor grows, the fractal dimension approaches $d - 1$. We can understand this fractal-dimension approach by recognizing that no population growth is purely fractal; it has a minimal length scale. As the population expands, the local curvature will flatten relative to the overall system size. Thus, the system will begin to appear more regular. The relative speed of the system's approach to seemingly regular surface growth might thus indicate the migratory nature of the underlying cell population, arising from a tumor's genetic (in)stability and energetic demand \cite{kimmel:CR:2020}.

Our framework also has implications for the spatially-extended reaction-diffusion models commonly used to describe tumor invasion. The Fisher-KPP equation and its variants, including the Porous-Fisher model with nonlinear diffusion and various power-law formulations, typically assume logistic growth kinetics in the reaction term \cite{simpson:PRSA:2024}. However, the appropriate form of density-dependence depends on the underlying spatial dynamics. Our results suggest that the well-mixed logistic term is justified only when cell migration is sufficiently rapid relative to proliferation; in low-migration regimes, growth is governed by geometric surface effects rather than mean-field kinetics, and generalized logistic or Gompertzian terms may be more appropriate. This distinction matters for model calibration: fitting standard logistic reaction-diffusion models to data from slowly migrating cell populations can yield good apparent fits but biased estimates of intrinsic growth rates. Our framework provides guidance on when logistic-based inference is reliable and when alternative growth laws should be considered.

A limitation of our approach is translation: clinical tumors are a diverse population of genetically and phenotypically heterogeneous cells with spatiotemporal variations in environmental resources, making it difficult to assess the nature of cell competition in patients. Our work focuses on 2D environments. In three-dimensional settings, diffusion-limited nutrient availability can impose additional constraints on proliferation, particularly in the tumor interior, such that the initial seeding density may have a different impact \cite{browning:eLife:2021}, possibly reinforcing surface-dominated growth patterns, though driven by metabolic limitation rather than contact inhibition alone. Also, our framework does not account for cell-cell adhesion or cohesion \cite{christgen:MBS:2025}, which could result in decreased migration rates and an increased likelihood of neighboring cells clumping together, thereby increasing patchiness in cell density.

In conclusion, our theory offers a possible explanation for the many mean-field laws that can adequately capture tumor growth dynamics. We do so by unifying density-dependent birth events via contact inhibition. In particular, our procedure provides a mechanistic underpinning for the widely observed patterns of Gompertzian growth driven by contact inhibition. This theoretical connection implies that complex cancer phenomena, such as tumor growth and dissemination, can be grounded in a few biophysical principles.


\section*{Resource Availability}

\subsection*{Data and code availability}
All code and data used in this manuscript are publicly available at \\\url{https://github.com/MathOnco/Contact-Inhibition}.

The agent-based model was implemented using the Hybrid Automata Library \cite{bravo:PCB:2020}. Data fitting was performed using Wolfram Mathematica (versions 12.0 and 14.1). Any additional information required to reanalyze the data reported in this paper is available from the lead contact upon request.


\section*{Acknowledgments}

We thank Philip Gerlee and the Integrated Mathematical Oncology Department members for helpful comments and discussions. This study was supported by the Richard O. Jacobson Foundation, Moffitt Cancer Center Evolutionary Therapy Center of Excellence, William G. 'Bill' Bankhead Jr and David Coley Cancer Research Program (20B06), National Cancer Institute (P30-CA076292 and U54-CA193489), and USAMRAA (KC180036). PMA is supported by the DFG Heisenberg program (grant no.~525136051).


\section*{Author Contributions}

G.J.K. and P.M.A. conceived of the model and designed the study. G.J.K., J.W., and P.M.A. carried out the analytical calculations. M.D. carried out experimental design, execution, and analysis. S.M. and J.W. carried out computational modeling analysis. G.J.K. and P.M.A. carried out statistical data analysis. A.R.A.A., J.W., and P.M.A. coordinated and supervised the study. G.J.K., S.M., M.D., A.T., J.W., and P.M.A. wrote the manuscript. All authors gave final approval for publication and agree to be held accountable for the work performed therein.


\section*{Declaration of Interests}

G.J.K. is an employee of Jacobs Levy Equity Management, Florham Park, NJ. P.M.A. declares funding from KITE (San Diego, CA) for research unrelated to this study and consultancy fees from CRISPR Therapeutics (Cambridge, MA) for unrelated work. 

\section*{Declaration of generative AI and AI-assisted technologies}

During the preparation of this work, the authors used Grammarly in order to improve grammar and writing style. After using this tool, the authors reviewed and edited the content as needed, and they take full responsibility for the manuscript's content.

\newpage

\onehalfspacing

\renewcommand\thefigure{S\arabic{figure}}
\setcounter{figure}{0} 
\renewcommand\thetable{S\arabic{table}}
\setcounter{table}{0} 
\setcounter{section}{0}

\section*{Supplementary Material: Mean-Field Equation}

\renewcommand{\theequation}{S.\arabic{equation}}

\vspace{-0.5cm}

Here, we sketch how the mean-field equation for the density $u=n/l$ can be derived from the master equation for the number of cells $n$ in a system of size $l$. Assume that the dynamics of $n=0,2,\dots,l-1,l$ follow a continuous-time birth-death Markov process. Let $P_n(t)$ be the probability density associated with finding the system with $n$ cells at time $t$ (typically given that it was in state $n_0$ at time 0, but we here omit this distinction for convenience and write $P_{n}(t)=P(n,t | n_0,t)$, as we do not consider a Kolmogorov backward equation). The following master equation describes the evolution of $P_n(t)$:
\begin{equation}
    \diff{P_n(t)}{t} = \lambda_{n-1} (n-1)P_{n-1}(t) + \delta_{n+1} (n+1)P_{n+1}(t)-(\lambda_n+\delta_n)\,n\,P_n(t) \label{master_eqn}
\end{equation}
where $\lambda_n$ denotes the birth rate and, $\delta_n$ denotes the population death. This master equation describes the stochastic dynamics in a discrete state space and continuous time. In a small time interval $dt$, the probabilities of jumping from $n$ to $n+1$ or $n-1$ can be written as $w_{n\to n+1}=\lambda_n\,dt$ and $w_{n\to n-1}=\delta_n\,dt$, and all other transions exept $w_{n\to n}$ occur with probability 0. 

The text books by Allen \cite{Allen:book:2003} and Gardiner \cite{gardiner:book:2004} show that the probaility density function of the continuous variable $u=n/l$ (where $l$ is large) is governed by the forward Kolmogorov equation
\begin{align}
	\frac{\partial\,p(u,t)}{\partial t} = -\frac{\partial}{\partial u}\left[ a(u)\,p(u,t) \right]+\frac{1}{2}\frac{\partial^2}{\partial u^2}\left[ b(u)\,p(u,t) \right] \label{forward_Kolmogorov}
\end{align}
where $a(u)$ is the drift coefficient and $b(u)$ is the diffusion coeffficient. In the deterministic limit, the deterministic analog of $u$ follows the ordinary differential equation 
\begin{align}
	\frac{\partial u}{\partial t} = a(u) \label{Liouville01}
\end{align}
with the initial condition $p(u,t|u_0,0)=\updelta(u-u_0)$ ($\updelta()$ is the Dirac delta function). The drift and diffusion coefficients can be derived from the process's central moments via the Kramers-Moyal expansion, which requires calculating the central moments \cite{Allen:book:2003,kampen:book:1997}. 

For the discrete system described by the master equation (\ref{master_eqn}), the central moments are
\begin{align}
	\mathbb E\left[ \left( n(t+dt)-n(t) \right)^k | n(t)=i \right] = \sum_{j=0}^{l}\left( j-i \right)^k\,w_{i\to j}. \label{CentralMoments01}
\end{align}
Now we are interested in rescaling of the variable $n$ and time $t$ by an appropriately chosen parameter, such that the limit in which this parameter vanishes yields the drift and diffusion coefficients from the central moments. A prudent choice is to scale with system size $l$: $m=n/l$ and $\tau=t/l$. Remembering that the transition probabilities $w_{i\to j}$ scale lineraly this $t$, which leads to
\begin{align}
	\mathbb E\left[ \left( m(\tau+d\tau)-m(\tau) \right)^k | n(\tau)=u \right] = \frac{l}{l^k}\sum_{j=0}^{l}\left( j-i \right)^k\,w_{i\to j}. \label{CentralMoments02}
\end{align}
Now we can use Equation (\ref{master_eqn}) and the definition of the $w_{i\to j}$ to work out the sum on the right-hand side:
\begin{align}
	\mathbb E\left[ \left( m(\tau+d\tau)-m(\tau) \right)^k | n(\tau)=u \right] = \frac{\lambda_m+(-1)^k\,\delta_m}{l^{k-1}}d\tau. \label{CentralMoments03}
\end{align} 
Then, the drift and diffusion coefficients follow from these central moments in the large system size limit as
\begin{align}
	a(u) &= \lim_{d\tau\to0}\frac{\mathbb E\left[ \left( m(\tau+d\tau)-m(\tau) \right) | m(\tau)=u \right]}{d\tau}, \label{DriftCoeff01}\\
	b(u) &= \lim_{d\tau\to0}\frac{\mathbb E\left[ \left( m(\tau+d\tau)-m(\tau) \right)^2 | m(\tau)=u \right]}{d\tau}, \label{DiffCoeff01}
\end{align} 
thus
\begin{align}
	a(u) &=  \left[ \lambda_m-\delta_m \right]_{m=u}, \label{DriftCoeff02}\\
	b(u) &=  \left[ \frac{\lambda_m+\delta_m}{l}\right]_{m=u}. \label{DiffCoeff02}
\end{align} 
Hence, the deterministic law for the density $u$ is given by 
\begin{align}
	\frac{\partial u}{\partial t} = \lambda_u-\delta_u, \label{Liouville01}
\end{align} 
and the diffusion coefficient vanishes with $l \to \infty$. 

\newpage

\section*{Supplementary Figures}

\makeatletter 
\renewcommand{\thefigure}{S\@arabic\c@figure}
\makeatother

\begin{figure}[b]
\begin{center}
\includegraphics[width=0.95\textwidth]{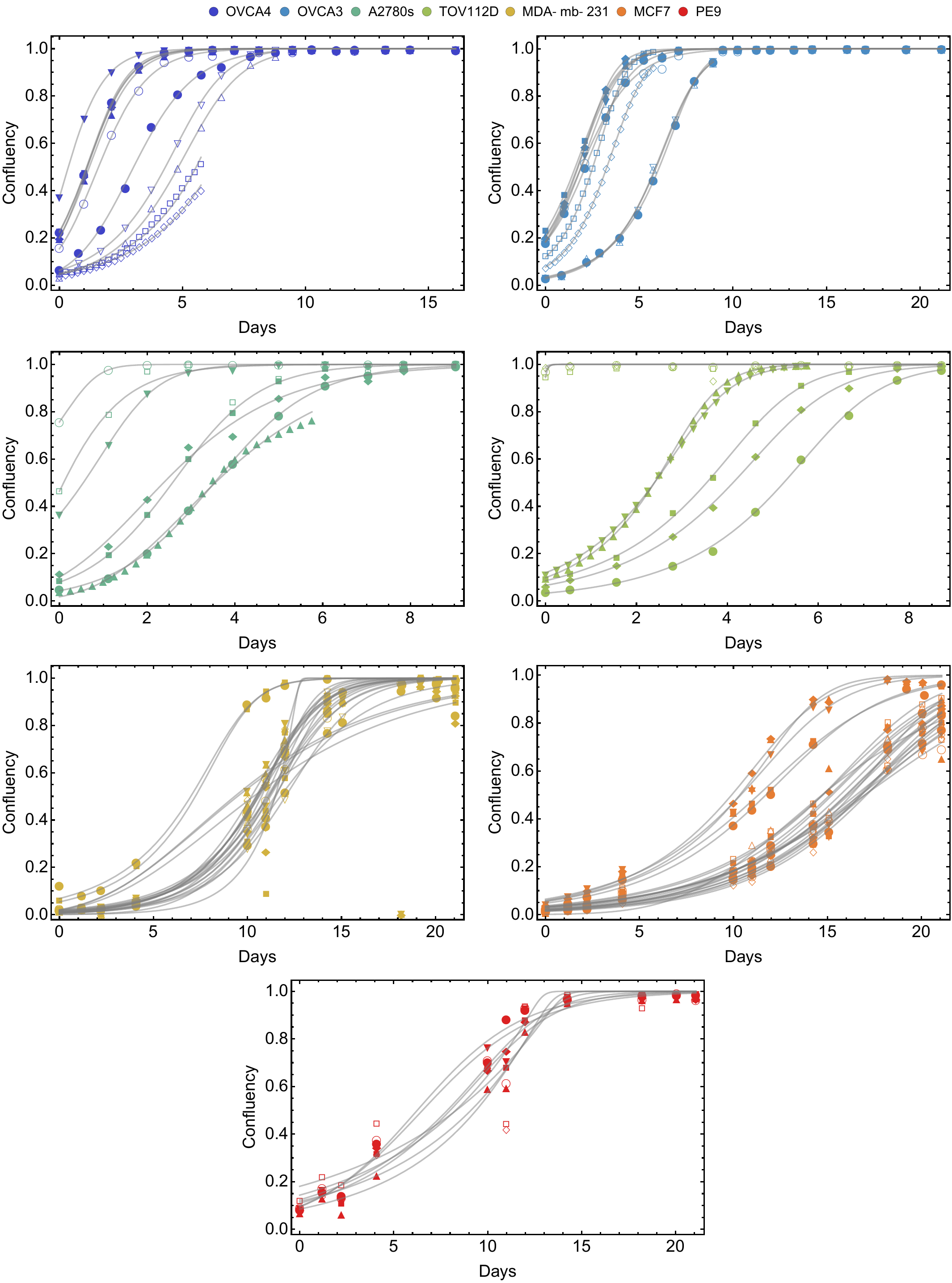}
\caption{
\textbf{
Fitting a generalized logistic growth law to longitudinal {\em in vitro} growth data of the seven cell lines (see Methods).
} 
}
\label{fig:SuppFig01}
\end{center}\end{figure}

\begin{figure}[b]
\begin{center}
\includegraphics[width=0.95\textwidth]{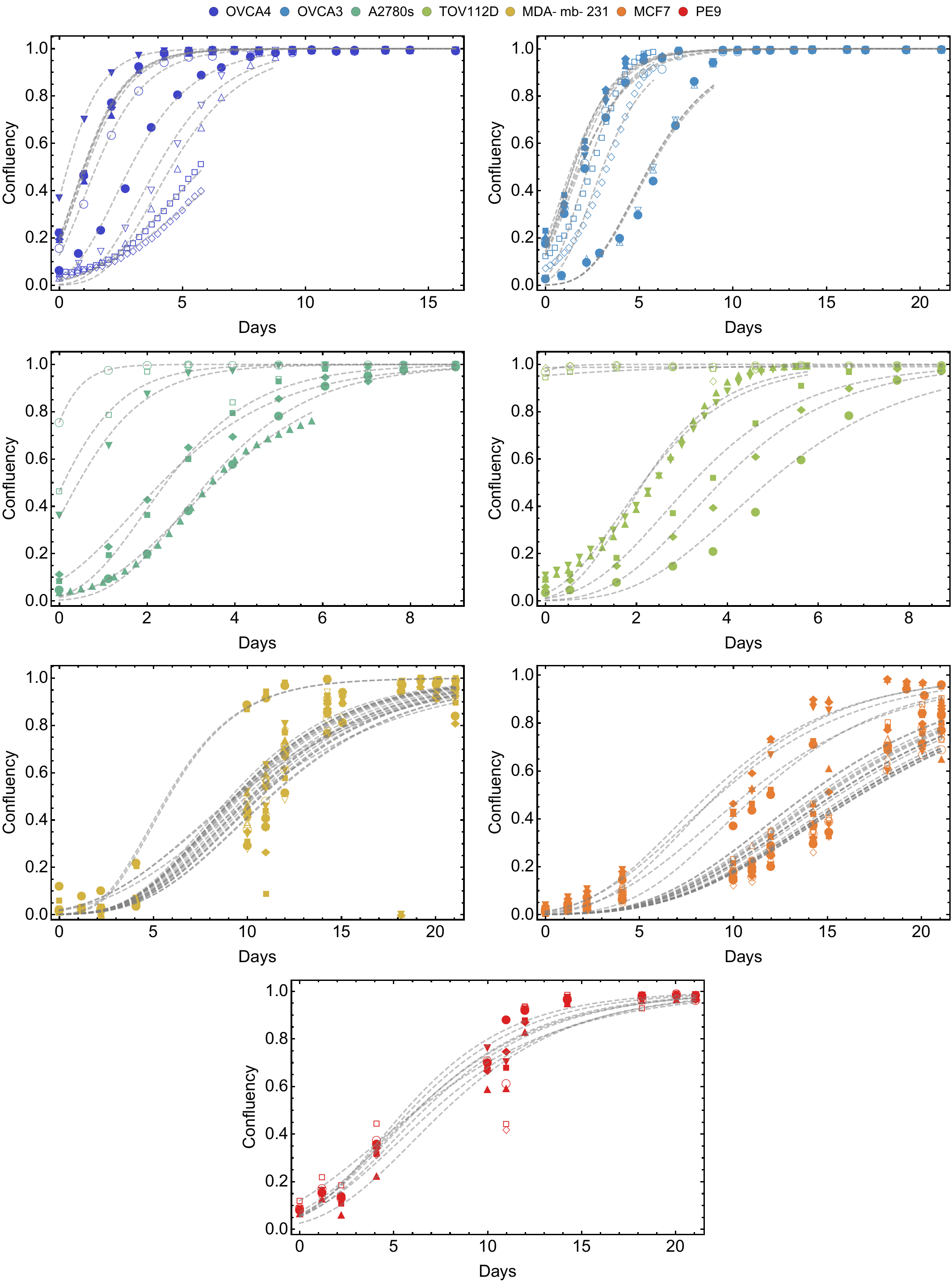}
\caption{
\textbf{
Fitting a Gompertz growth law to longitudinal {\em in vitro} growth data of the seven cell lines (see Methods).
}
}
\label{fig:SuppFig02}
\end{center}\end{figure}

\begin{figure}[b]
\begin{center}
\includegraphics[width=0.95\textwidth]{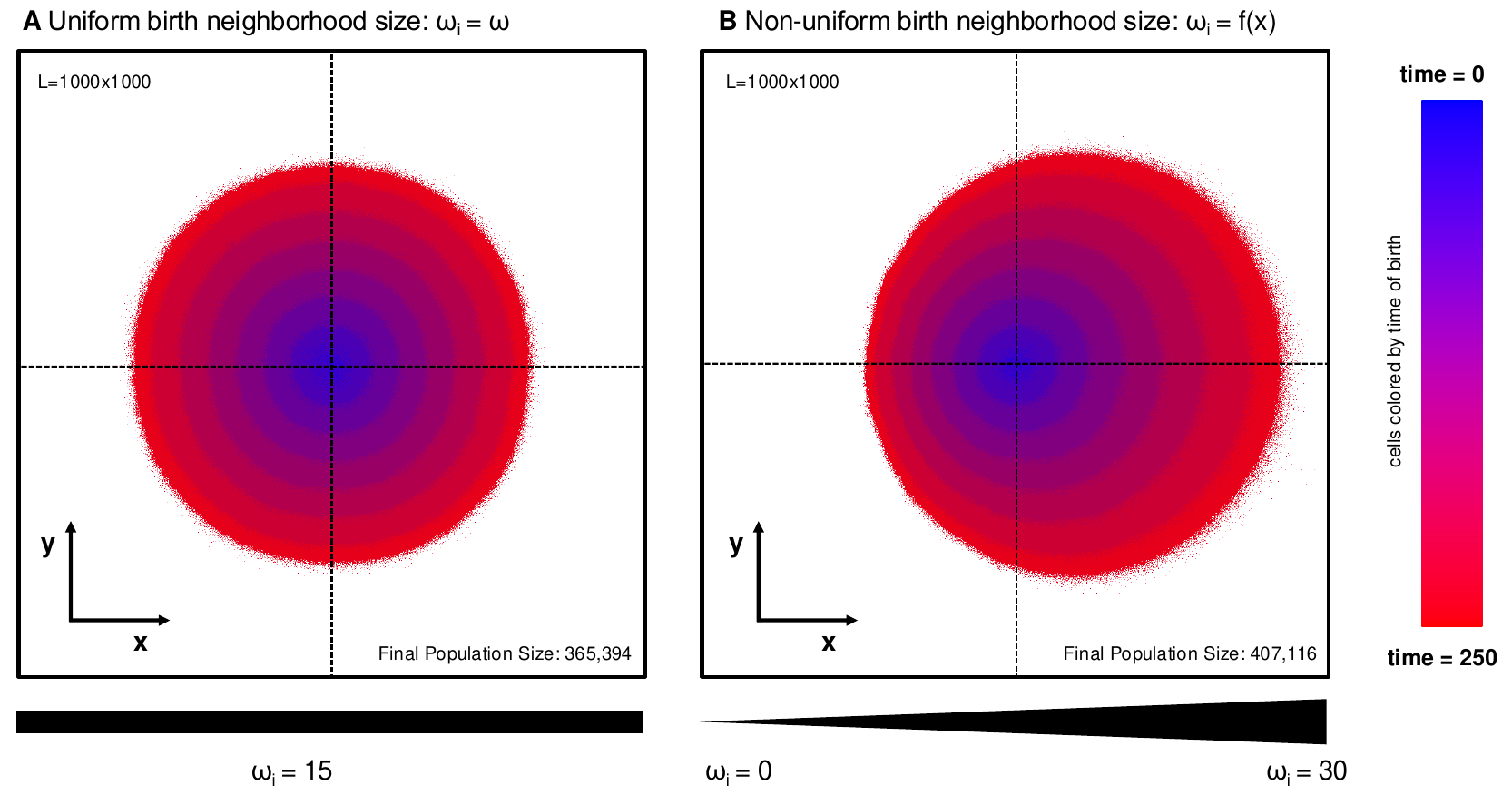}
\caption{
\textbf{ Uniform versus non-uniform birth neighborhood size in space. 
} (A) A single stochastic simulation ($l=10^6$; $b=0.1$) with a constant birth neighborhood size ($\omega_i=15$) seeded with a single cell at the center of the domain. (B) An identically parameterized stochastic simulation as in (A), but with non-uniform, varying birth neighborhood size. On the left-hand side of the domain $\omega_i=0$, while increasing linearly such that the right-hand side $\omega_i=30$. Despite having the same average neighborhood size throughout the domain, the non-uniform simulation grows to a larger population size, and is skewed toward the right-hand side. Dashed lines indicate the domain center, to guide the eye.
}
\label{fig:SuppFig03}
\end{center}\end{figure}

\begin{figure}[b]
\begin{center}
\includegraphics[width=0.95\textwidth]{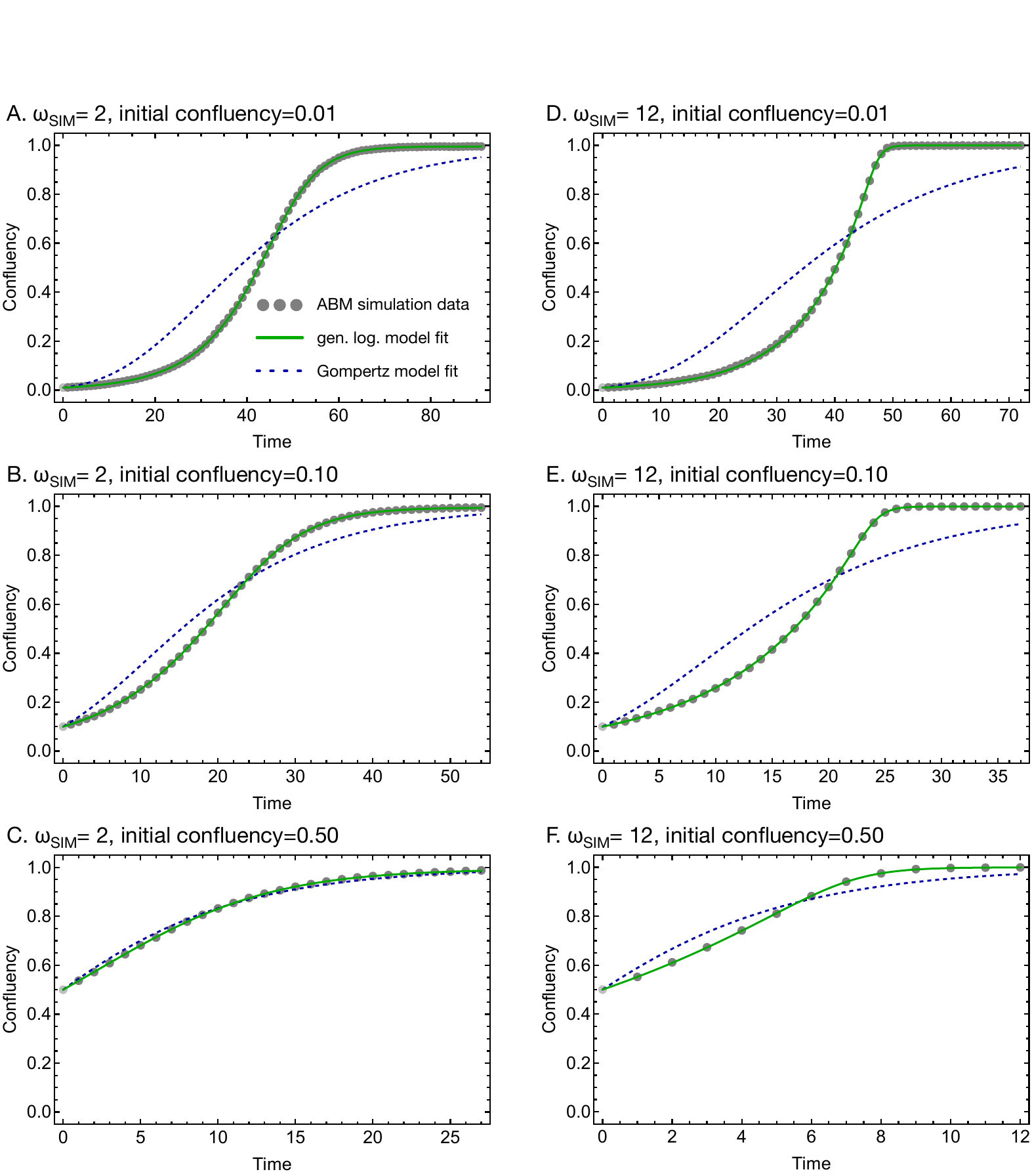}
\caption{
\textbf{ Example fits of generalized logistic (gen.~log.) and Gompertz models to agent-based model simulation data.
} 
Generalized logistic model (solid lines) and Gompertz model (dashed lines) fits do different simulated agent-based (ABM) models, across different initial conditions (confluencies) and for different implemented birth-neighborhood values $\omega_{\text SIM}$. 
}
\label{fig:SuppFig04}
\end{center}\end{figure}

\end{document}